\documentclass[a4paper,11pt]{article}
\usepackage[utf8]{inputenc}
 \usepackage{fouriernc}

\usepackage{authblk}
\usepackage{microtype}
\usepackage{amsmath,mathtools}
\usepackage{amsfonts}
\usepackage{amssymb}
\usepackage{mathrsfs}
\usepackage{bbm}
\usepackage[english]{babel} 
\usepackage[compress]{cite}
\usepackage{eufrak}
\usepackage{yfonts}
\usepackage{bmpsize-base}
\usepackage{fullpage}
\usepackage{setspace}
\usepackage{booktabs}
\usepackage{eso-pic}
\usepackage{shapepar}
\usepackage{enumitem}
\usepackage{braket}
\usepackage{graphicx}
\usepackage{epstopdf}
\usepackage{caption}
\usepackage{float}
\usepackage{subfigure}
\usepackage{bm}
\usepackage{makeidx}
\usepackage{placeins}
\usepackage{listings} 
\usepackage{color} 
\usepackage{titlepic}
\usepackage{bbm}
\usepackage{mathrsfs}	
\usepackage{soul}
\usepackage{color}
\usepackage{scalerel}
\usepackage{framed,lipsum}
\definecolor{gray}{rgb}{92.,92.,92.}
\newcommand {\Cdot}{\vcenter{\hbox{\scalebox{1.6}{.}}}}
\title{The full replica symmetry breaking in the Ising spin glass on random regular graph}

\date{}
\author{Francesco Concetti}
\affil{\textit{Dipartimento di Fisica, Sapienza Università di Roma, Piazzale Aldo Moro 2, I-00185 Rome, Italy}}
\begin{document}

\maketitle
\begin{abstract}
In this paper, we extend the full replica symmetry breaking scheme to the Ising spin glass on a random regular graph. We propose a new martingale approach, that overcomes the limits of the Parisi-Mézard cavity method, providing a well-defined formulation of the full replica symmetry breaking problem in random regular graphs. Finally, we define the order parameters of the system and get a set of self-consistency equations for the order parameters and the free energy. We face up the problem only from a technical point of view: the physical meaning of this approach and the quantitative evaluation of the solution of the self-consistency equations will be discussed in next works. 

\end{abstract}
\section{Introduction}
The mean-field theory of spin glasses has attracted a considerable interest in the last forty years, as a promising theory to describe the statistical mechanics of glassiness and disorder systems.

The Sherrington-Kirkpatrick \cite{SK1,SK2} model has represented a fruitful source of insights in this field.
The complete solution was derived by Parisi via the replica method \cite{EA}, with the introduction of a clever Replica-Symmetry-Breaking ansatz (RSB) \cite{Par1_0,Par1_1,Par1_2}.
The Parisi RSB ansatz was further clarified by Mézard, Parisi and Virasoro \cite{VPM} as related to the decomposition of the Gibbs state in a mixture of a large number of pure equilibrium states, recognizing the Parisi RSB order parameters \cite{Par2}, i.e. the overlaps between replicas, as the distribution of the mutual overlap between two equilibrium states on the phase space.

The replica method was successfully used in many problems on fully-connected networks \cite{VPM,Nishimori}. 
Various models show different patterns of RSB, depending on the way the states are “distant” to each other. 
\begin{itemize}
\item The overlaps between different states can take (almost surely) only two different values. In this case, we speak about “one-step replica symmetry breaking”($1$-RSB) solution. The states are scattered randomly in the phase space and correspond to stable well-defined minima (genuine minima) of the free energy landscape \cite{REM, pSpinCri}. 
\item The overlaps can take a discrete number $r+1 \in \mathbb{N}$ of values in the interval $[q_m : q_M]$. In this case, we speak about “$r$-step replica symmetry breaking”($r$-RSB) solution. The equilibrium states exhibit a hierarchal structure, where clusters of states with a given mutual overlap are grouped in a progressively wider level of clusters with a progressively lower overlap, for $r$ levels \cite{ViraMezUltra,VPM}. Each state enters in the Gibbs decomposition with a random weight, which is generated according to the Derrida’s REM and GREM calculations \cite{DerridaGREM, ParisiMezardGREM}. Such solutions can be considered as an iterative composition of $1$-RSB solutions. As far I know this situation is very rare.
\item The overlaps among states can take all possible values in the interval $[q_m : q_M]$. In this case, we speak about “full replica symmetry breaking”(full-RSB) and it can be considered as a $r\to \infty$ of the preceding case. The equilibrium states exhibit a continuous fractal clustering and the random weights are configurations of a Ruelle random probability cascade \cite{Ruelle,ASS,Arguin}, that provides a continuous extension of GREM. It is worth stressing that, unlike the preceding case, the equilibrium states can be arbitrarily close and the barriers between the states may be very small. The minima are marginal, with many flat directions ( infinite in the thermodynamic limit) \cite{DeDoKondor}.
\end{itemize}

Most of the results of these results have been reproduced using a probabilistic iterative approach, the cavity method, which avoids the mathematical weirdnesses of the replica method \cite{VPM, ASS}. The replica jargon, however, is used in spin glass theory, regardless the approach considered: actually, we speak about RSB if the system exhibits many pure states, organized according to one of the schemes described before.

The rigorous proof of the Parisi os the SK model solution was derived by Talagrand \cite{Tala}, with the Guerra's interpolation scheme, that allows a rigorous handling of the replica symmetry breaking\cite{Guerra}. The ultrametricity of the states was proved soon after by Panchenko \cite{PanchenkoUltra,PanchenkoUltra2}, in relation to the notion of stochastic stability \cite{GhirlandaGuerra}.

If and how the RSB scheme applies also in non-fully-connected systems is still debated, in spite of recent results.

We concentrate our attention on spin glasses defined on a random regular graph (RRG), with finite connectivity \cite{Bellobas}.

The interest in sparse graph models is motivated by the fact that they represent a more realistic class of mean-field models, including the notion of neighborhood which is absent in the infinite range case. They attract a big interest also in computer science, since many random optimization problems turn out to have a finite connectivity structure.

The $1$-RSB scheme was successfully extended to the Ising spin glass on sparse graphs by Parisi and Mézard (PM) with the cavity method \cite{ParMezRRG1,ParMezRRG2,ParMezRRG4,MonaMez}, improving the Bethe–Peierls method in order to deal with many equilibrium states. The approach can be easily generalized to the case with $r$ steps of replica symmetry breaking, imposing the GREM scheme for the states distribution and the ultrametricity as in the fully-connected case \cite{Panchenko2015,Panchenko2016}. 

It was proved, via interpolation arguments, that this approach should provide a rigorous lower bound for the free energy \cite{FranzLeone,FranzLeone2}.

The $1$-RSB PM cavity method has been very successful even now, since, in certain regimes, this approach provides an algorithmic solution of random satisfiability problems with finite connectivity\cite{ParMezRRG2, MonaMez}. 

Unfortunately, in the PM cavity method approach, the $r$-RSB order parameter is a distribution of $(r-1)$-RSB order parameters. Since the replica symmetric solution already involves an order parameter which is the local fields distribution, going to a $1$- RSB solution the replica order parameter becomes now the probability distribution over the space of local field distributions and the $2$-RSB free energy order parameter is a distribution of distributions of distributions \cite{ParMezRRG1}. As a consequence, the cavity method turns out to be inadequate to achieve a full-RSB theory for RRG spin glasses, indeed the $r\to \infty$ limit of the order parameter has no mathematical meaning. Also, the high levels of replica symmetry breaking are actually numerically intractable.

Actually, the $1$-RSB cavity method is commonly used also to describes models where the equilibrium states are not independent and well-defined minima ($1$-RSB approximation). This approximation, however, cannot grab the marginality of the states and then completely misses the right evaluation of such quantities that have very different properties in the marginally stable phase, as the spectrum of small oscillations, nonlinear susceptibilities and so on \cite{ParisiMarginal}. This limitedness entails the impossibility to describe, with the cavity method, a glassy phase with many marginal equilibrium states. 

 In this paper, we propose an alternative approach, that overcomes the difficulties presented by the cavity method and allows to get a mathematical formulation of the full-RSB free energy for Ising spin glasses on RRG. 

We manage the progressive branching of the clusters of states with a martingale approach \cite{YoRev}, improving the idea suggested in \cite{ParisiMarginal}. We reduce the computation to a series of variational problems, where variational parameters are martingales.
The order parameter, then, is not a deterministic distribution, as in the cavity method, but it is a stochastic process. 

The paper is organized as follows.

In the next section, we present the model. In the third section, we present the cavity variational free energy with the $r$-RSB ansatz that extends to RRG the hierarchical ROSt formalism proposed by Azienman, Sims and Starr for the SK model \cite{ASS}. Using the properties of GREM and standard technics of statistical mechanics, we rewrite the $r$-RSB as the $r$th iteration of a discrete time recursive low, starting from the RS free-energy and going down along the various levels of clustering of the equilibrium states. The previous recursion low is then reconsidered using a discrete time martingale approach. 

In the last section, the formalism can be extended to continuous-time, obtaining an auxiliary variational representation à la Boué and Dupuis \cite{BoueDepuis} for the full-RSB variational free energy functional. Suitable full-RSB order parameters are defined and the variational free energy functional is finally derived as a generalization of the Chen and Auffinger representation of the Parisi free energy for SK model \cite{ChenAuf}. The self-consistency equations for the order parameters are then derived from the first variation of the free energy functional. 

Whilst a rigorous proof is still lacking, it is generally believed that the full-RSB ansatz is correct for Ising spin glasses on sparse graphs. If it is the case, the equations presented in this work should provide the right solution for this model. In any case, the mathematical formulation of the full-RSB theory provides a basic groundwork for complete mathematical proof about RSB on non-fully-connected models, extending the recent results on discrete RSB \cite{Panchenko2015}.

The aim of the paper is to provide a clear formal definition and a non-ambiguous mathematical setting for the full-RSB problem for spin glass models in random regular graphs. We will tackle this problem only from a technical point of view. The physical interpretation of our approach will be discussed further in next works.

\section{The model}
We consider a system of $N$ Isisng spins $\bm{\sigma}:=(\sigma_1,\sigma_2,\cdots,\sigma_N)\in\{-1,1\}^N$ with the Hamiltonian:
\begin{equation}
\label{hamiltonian}
H[J,\mathcal{G}_c,\bm{\sigma}]=\sum_{\braket{ij}_{\mathcal{G}_c}}J_{i,j}\sigma_i\sigma_j
\end{equation}
The sum runs over the edges of a random regular graph $\mathcal{G}_c$, with connectivity $c$ ($RRG_c$). The couplings $\{J_{i,j}\}$ are independent $\{-1,1\}-$valued random variables. 
Our aim is to compute the free energy density, defined as:
\begin{equation}
\label{aim}
F_{J,\mathcal{G}_c}=\lim_{N\to \infty}-\frac{\log Z_{N,J,\mathcal{G}_c}}{\beta N}=\lim_{N\to \infty}-\frac{1}{\beta N}\,\log\sum_{\bm{\sigma}\in \{-1,1\}^N}e^{-\beta H[J,\mathcal{G}_c,\bm{\sigma}]}\,,
\end{equation}
where $Z_{N,J,\mathcal{G}_c}$ is the partition function on an N spins system. The couplings and the graph constitute the quenched disorder of the system.

 It can be shown that, in the thermodynamic limit, the free energy does not depend on the realization of the quenched disorder with probability one. As a consequence, we will concentrate on the computation of the average free energy:
\begin{equation}
\label{aim}
F=\lim_{N\to \infty}-\frac{1}{\beta N}\overline{\log \,Z_{N,J,\mathcal{G}_c}\,}
\end{equation}
where $\overline{\,\cdot\,}$ denotes the average over the random couplings and the realizations of the random graph.

Because of the randomness of the edges, for large $N$ and $c> 2$, the typical size of a loop is of order $\sim \log N$, so the probability to have finite loops vanishes in the $N\to \infty$ limit \cite{Bellobas}. As consequence,
random regular graphs are locally isomorphic to a tree: such models are exactly solvable in mean-field theory. Large loops, however, can induce frustration, so the system exhibits a glassy behavior at low temperature.

At high temperature, the system exhibits a single pure state (the paramagnetic one), and the free energy can be exactly solved by the Bethe-Peierls (BP) approximation. However the BP approach turns out to be wrong at low temperature, where the Gibbs state is no more a clustering state, i.e. the correlation function does not vanish exponentially with the distance, invalidating the BP assumptions \cite{ParMezRRG1}. A more sophisticated cavity approach must be considered, dealing with the presence of many clustering states. The BP approximation still holds within each state separately, whilst the average over the states generate non-vanishing correlation functions for the Gibbs state \cite{ParMezRRG1,ParisiMarginal}.

In the $c \to \infty$ limit, keeping $c\Braket{J^2} = 1$, the free energy becomes independent of the probability distribution of the $J$ and we obtain the same free energy of the SK model.

\section{The finite replica symmetry breaking problem}
In this section, we describe the discrete RSB scheme for this model. 

In the first subsection, we define the $r$-RSB cavity free energy functional for sparse graphs. We recall the notion of discrete Ruelle random probability cascade, or GREM \cite{DerridaGREM, Ruelle, ASS, Arguin}, and generalize the Parisi RSB ansatz for diluted models \cite{Panchenko2015,Panchenko2016}.

In the second subsection, we recast the progressive steps of replica symmetry breaking in a discrete time recursive map, that generalizes the Parisi replica computation for the SK models \cite{Par1_0,Par1_1,Par1_2, VPM}. 

In the last subsection, we derive a new variational representation of the $r$-RSB free energy, using a progressive iteration of the Gibbs variational principle. 
\subsection{The finite RSB free energy functional}
\label{finitersbpara}
Let us assume that the system has many equilibrium states, that are labeled by an index $\bm{\alpha}$. The cavity spins are uncorrelated within a given state $\bm{\alpha}$, leading to a factorized cavity spins distribution, that depends on the label $\bm{\alpha}$. Since each spin $\sigma_i$, with $1\leq i \leq N$, can take only two values, the cavity probability distribution, for a given state $\bm{\alpha}$, depends only on the cavity magnetization $m_{i|\bm{\alpha}}$ or, equivalently, on the cavity field $h_{i|\bm{\alpha}}=1/\beta\,\,\text{atanh}\, m_{i|\bm{\alpha}}$. The cavity fields depend on the random couplings, so they are also random quantities and their distribution is not known a priori. The equilibrium free energy is finally given by the Gibbs state, that is a statistical mixture of the states $\bm{\alpha}$.

The cavity free energy functional is given by \cite{ParisiMarginal}
\begin{equation}
\label{cavityFE}
\Phi= \overline{\mathbb{E}\log\left(\frac{\sum_{\bm{\alpha}}\xi_{\bm{\alpha}}\Delta^{\text{(v)}}(\bm{J},\bm{h}_{\bm{\alpha}})}{\sum_{\bm{\alpha}}\xi_{\bm{\alpha}}}\right)}-\frac{c}{2}\overline{\mathbb{E} \log\left(\frac{\sum_{\bm{\alpha}}\xi_{\bm{\alpha}}\Delta^{\text{(e)}}(J_{1,2},h_{1|\bm{\alpha}},h_{2|\bm{\alpha}})}{\sum_{\bm{\alpha}}\xi_{\bm{\alpha}}}\right)}\,.
\end{equation}
Here $\bm{J}=(J_{0,1},J_{0,2},\cdots,J_{0,c})$ and $\bm{h}_{\bm{\alpha}}=(h_{1|\bm{\alpha}},h_{2|\bm{\alpha}},\cdots,h_{c|\bm{\alpha}})$ and the variables $\{\xi_{\bm{\alpha}}\}_{\bm{\alpha}}$ are the (non-normalized ) statistical weights of the states. All the $c+2$ couplings in the functional \eqref{cavityFE} are independent.

The functions $\Delta^{\text{(v)}}$ and $\Delta^{\text{(e)}}$ are defined as:

\begin{gather}
\label{deltaterms}
\Delta^{\text{(v)}}(\bm{J},\bm{h}_{\bm{\alpha}})=\cosh\big(\beta U_c(\bm{J},\bm{h}_{\bm{\alpha}})\,\big)\prod^c_{i=1}\frac{\cosh(\beta J_{0,i})}{\cosh\big(\,\beta u(J_{0,i},h_{i|\bm{\alpha}})\,\big)}\,,\\
\Delta^{\text{(e)}}(J_{1,2},h_{1|\bm{\alpha}},h_{2|\bm{\alpha}})=\cosh(\beta J_{1,2})\big(1+\tanh(\beta J_{1,2})\tanh(\beta h_{1|\bm{\alpha}})\tanh(\beta h_{2|\bm{\alpha}})\,\big)\,,
\end{gather}

with
\begin{gather}
\label{U}
u(J_{1,2},h_{2|\bm{\alpha}})=\frac{1}{\beta}\text{atanh}\big(\tanh(\beta J_{1,2})\tanh(\beta h_{2|\bm{\alpha}})\,\big)\,,\\
U_c(\bm{J},\bm{h}_{\bm{\alpha}})=\sum^c_{i=1}u(J_{0,i},h_{i|\bm{\alpha}})\,.
\end{gather}
The overline $\overline{\,\cdot\,}$ stands for the average over the quenched disorder and the expectation value $\mathbb{E}$ is over all the cavity fields and the random weights $\{\xi_{\bm{\alpha}}\}_{\bm{\alpha}}$ .

The contribution to the free energy \eqref{cavityFE} that depends on the function $\Delta^{\text{(v)}}$ is usually called \textit{vertex contribution}, whilst the contribution depending on $\Delta^{\text{(e)}}$ is the \textit{edge contribution}.

The equilibrium free energy is, formally given by:
\begin{equation}
\label{EqFreeEnRPC}
-\beta F=\min_{\mathcal{P}(\{\xi _{\bm{\alpha}}\}_{\bm{\alpha}},\{h_{i|\bm{\alpha}}\}_{i,\bm{\alpha}})}\Phi\,,
\end{equation}
where the supremum must be take over the set of all the possible probability distributions of the cavity fields $\{h_{i|\bm{\alpha}}\}_{i,\bm{\alpha}}$ and the random weight of the state $\{\xi _{\bm{\alpha}}\}_{\bm{\alpha}}$.
This set is huge and too general, then further assumptions are needed to face up the problem.

In the Parisi-Mézard RSB ansatz, the sum $\sum_{\bm{\alpha}}\,\cdot\,$ runs over the leaves of an infinitary rooted taxonomic tree and $\bm{\xi}:=\{\xi_{\bm{\alpha}}\}_{\bm{\alpha}}$ is a collection of positive random variables generated by a Ruelle random probability cascade defined along the tree; for each site $i$, the set $\{h_{i|\bm{\alpha}}\}_{\bm{\alpha}}$ is a random hierarchical population of fields generated along the same tree. Such hierarchical populations are independent for different site index $i$ and identical distributed. 

More specifically, the $r$-RSB ansatz, for a finite integer $r$, is defined as a generalization of the Aizenman-Sims-Starr (ASS) \cite{ASS}
 construction of the hierarchal Random Overlap Structucture (ROSt) for the SK model \cite{Panchenko2016,PanchenkoExchange}. 

Let $X$ be a non-decreasing sequence of $r+2$ numbers, for some $r\in\mathbb{N}$:
\begin{equation}
0=x_0 \leq x_1\leq \cdots\leq x_r \leq x_{r+1}=1\,.
\end{equation}
We first define a Poisson point process $\bm{\xi}^{(1)}:=\{\xi_{\alpha_1}^{(1)};\,\alpha_1 \in \mathbb{N}\}$ on $[0,\infty)$, with density given by $\rho(d\xi)=x_1 \xi^{-x_1-1} d\xi$; such a process is usually referred to as $REM_{x_1}$.

Next, for each $\alpha_1$, a $REM_{x_2}$ process $\bm{\xi}_{\alpha_1}^{(2)}:=\{\xi_{(\alpha_1,\alpha_2)}^{(2)};\,\alpha_2 \in \mathbb{N}\}$ is generated, independently for different values of $\alpha_1$. We then iterate the procedure: at the $n-$th level, up to $n=r$, independent realizations of the $REM_{x_n}$ process $\bm{\xi}_{(\alpha_1,\cdots,\alpha_{n-1})}^{(n)}:=\{\xi_{(\alpha_1,\cdots,\alpha_{n-1},\alpha_{n})}^{(n)};\,\alpha_{n} \in \mathbb{N}\}$ are generated for each of the distinct values of the multi-index $(\alpha_1,\alpha_2,\cdots,\alpha_{n-1})$ of the previous iteration.
Let also introduce the quantity $\xi_{\star}=1$.

Such structure defines an infinitary rooted taxonomic tree of depth $r$, with the vertex set given by
\begin{equation}
\mathcal{A}=\{\star\}\cup\mathbb{N}^1\cup\mathbb{N}^2\cup\cdots\cup\mathbb{N}^r
\end{equation}
with each vertex $(\alpha_1,\cdots,\alpha_{n-1})$ branching to the vertices $(\alpha_1,\cdots,\alpha_{n-1},\alpha_n)$, for all $\alpha_n\in \mathbb{N}$. We denote by $|\bm{\alpha}|$ the level, i.e. the lenght, of the multi-index $\bm{\alpha}\in\mathcal{A}$, with $|\star|=0$.

Each $\bm{\alpha}=(\alpha_1,\alpha_2,\cdots,\alpha_r)\in\mathbb{N}^r$, at the boundary, identifies a path along the tree, defined as:
\begin{equation}
\bm{\alpha}\mapsto p(\bm{\alpha})=\left\{\,\star,(\alpha_1),(\alpha_1,\alpha_2),\cdots,(\alpha_1,\alpha_2,\cdots,\alpha_r)\,\right\}.
\end{equation}
The vertex $\star$ is the starting point of all the paths. 

The $r-$step Ruelle random probability cascade, for the sequence $X$, or GREM$_X$, is then defined as the point process $\{\xi_{\bm{\alpha},r}\}_{\bm{\alpha}\in\mathbb{N}^r}$ such that:
\begin{equation}
\xi_{\bm{\alpha},r}=\prod_{\bm{\beta}\in p(\bm{\alpha})}\xi_{\bm{\beta}}^{(|\bm{\beta}|)}=\xi_{\star}^{(0)}\xi_{(\alpha_1)}^{(1)}\xi_{(\alpha_1,\alpha_2)}^{(2)}\cdots \xi_{(\alpha_1,\alpha_2,\cdots,\alpha_r)}^{(r)}.
\end{equation}

Note that a rigorous definition of the Ruelle probability cascade point process requires the reordering, for each level $0\leq k\leq r+1$, of the random variables, generated in $REM_{x_k}$, in a decreasing order \cite{DerridaGREM, Ruelle, ASS, Arguin, Panchenko2015}.

For any given site $i$, the population of cavity fields $\{h_{i|\bm{\alpha},r}\}_{\bm{\alpha}\in\mathbb{N}^r}$ is a random array, that is assumed to be independent of the random weights $\{\xi_{\bm{\alpha},r}\}_{\bm{\alpha}\in\mathbb{N}^r}$
 and \emph{hierarchical exchangeable}, i.e. the distribution is invariant under permutations that preserve the tree structure; such assumption is the key of the Parisi-Mézard ansatz \cite{ParMezRRG1,ParMezRRG2,ParMezRRG4,MonaMez,Panchenko2015,Panchenko2016,FranzLeone,FranzLeone2,ParisiMarginal} and it turns out to be exact, assuming the validity of the Ghirlanda Guerra identities \cite{PanchenkoExchange}. 

Furthermore, by general argument, we can safely argue that all the cavity fields have zero mean and are almost surely bounded:

\begin{equation}
\mathbb{E}[h_{i|\bm{\alpha},r}]=0
\end{equation}
\begin{equation}
\mathbb{E}[|h_{i|\bm{\alpha},r}|]<\infty\quad \text{for all allowed $i$ and $\bm{\alpha}$}\,.
\end{equation}

In the ASS hierarchal ROSt \cite{ASS}, the populations of cavity fields are generated by defining, independently for each site $i$, a set of independent Gaussian variables, labelled by the vertices of the taxonomic tree $\mathcal{A}$, and representing each cavity fields $h_{i|\bm{\alpha},r}$ by the sum over the Gaussian variables corresponding to the vertices of the path $p(\bm{\alpha})\subset\mathcal{A}$.

Gaussianity is too restrictive for the actual model and a more general distribution must be considered.

A more generic hierarchical exchangeable random array can always be represented by the hierarchical version of the Aldous-Hoover theorem, presented in \cite{Austin,AustinPanchenko}.

As in the ASS work, for any given index $i$, let $\{W^{(|\bm{\alpha}|)}_{i|\bm{\alpha}}\}_{\bm{\alpha}\in \mathcal{A}}$ be a collection of independent and identical normal distributed random variables\footnote{In the original works by Austin and Panchenko \cite{Austin,AustinPanchenko}, a random array is generated by a function of unifom random variables on $[0,1]$. A uniform random variable, however, can be generated in distribution as a function of a Gaussian variable, than the representation presented here is equivalent to Austin and Panchenko representation. }, labelled by the vertices of the taxonomic tree $\mathcal{A}$, and consider a measurable function $h:\mathbb{R}^{r+1}\to \mathbb{R}$, which we will refer to as \emph{cavity field functional}. The cavity field population, at the site $i$, can be generated by presenting each cavity fields $h_{i|\bm{\alpha},r}$ as follow:
\begin{equation}
\label{AustinRep}
h_{i|\bm{\alpha},r}=h_r\big(\,\big\{\,W^{(|\bm{\beta}|)}_{i|\bm{\beta}}\big\}_{\bm{\beta}\in p(\bm{\alpha})}\,\,\big)=h_r\big(\,W^{(0)}_{i|\star},W^{(1)}_{i|(\alpha_1)},W^{(2)}_{i|(\alpha_1,\alpha_2)},\cdots,W^{(r)}_{i|(\alpha_1,\alpha_2,\cdots,\alpha_r)}\,\big)\,.
\end{equation}
 
The variable $W^{(0)}_{i|\star}$ is the root random variable of the site $i$ and it is shared amongst all the $\bm{\alpha}$s. The collections $\big\{\,W^{(|\bm{\beta}|)}_{i|\bm{\beta}}\big\}_{\bm{\beta}\in p(\bm{\alpha})}$ are independent for different sites $i$.
 
Taking the average over all the random quantities, the functional $\Phi$ will depends only on the sequence $X$ and on the cavity field functional $h$. The equilibrium free energy is given by the extremizing the functional $\Phi$ with respect to such two parameters. 

The cavity field functional is the actual order parameter of the model and encodes entirely the Parisi-Mézard ansatz for the cavity fields distributions inside the pure states \cite{Panchenko2016}, as shown in the appendix for the $1-$RSB case \ref{appendix}.

The cavity field functional turns out to be a handier order parameter than the recursive tower of distributions on the set of distributions presented in the Parisi-Mézard original works \cite{ParMezRRG1,ParMezRRG2,ParMezRRG4,MonaMez} and can be easily extended to the full-RSB case. It is worth noting, however, that the representation \eqref{AustinRep} is quite redundant, indeed there are many choices of the function $h$ that will produce the same array in distribution \cite{Panchenko2016}. This redundancy constitutes a gauge symmetry and the distribution of the cavity field population is the gauge invariant order parameter.
 
If the cavity field functional is linear, the representation \eqref{AustinRep} recovers the ASS hierarchal ROSt scheme. As discussed in the next subsection, in case of linearity, or additive separability at least, the free energy can be represented as the solution of a proper partial (integro-)differential equation, like the Parisi solution of the SK model \cite{Par1_1,VPM}.
Additive separability, however, fails to fit the results emerging at $1-$RSB levels \cite{ParMezRRG1,ParMezRRG2,ParMezRRG4,MonaMez}. 

As we shall see below, the martingale approach to the cavity method allows dealing with a generic cavity field functional $h$, leading to a well-defined $full-$RSB theory, with an explicit definition of the order parameter, an explicit representation of the functional \eqref{cavityFE} and a proper self-consistency mean-field equation.

Note that, for a fixed state $\bm{\alpha}$ and $i$, the distribution of the $r+1$ random variables $\{W^{(|\bm{\beta}|)}_{i|\bm{\beta}}\}_{\bm{\beta}\in p(\bm{\alpha})}$ does not depend explicitly on the multi-index $\bm{\alpha}$, and, in the following, we will drop it away without ambiguities:
\begin{equation}
\label{realizationfixed}
\{W^{(|\bm{\beta}|)}_{i|\bm{\beta}}\}_{\bm{\beta}\in p(\bm{\alpha})}\longrightarrow \{W^{(m)}_i\}_{m\leq r}:= \,\,\{W_i^{(0)},W_i^{(1)},W_i^{(2)},\cdots,W_i^{(r)}\}\,,\\
\end{equation}

We will also indicate with $\{W^{(m)}_i\}_{m\leq n}$ the set of all the variables $W_i^{(m)}$, along a given path, from the level $m=0$, to the level $m=n\leq r$:
\begin{equation}
\begin{gathered}
\label{realizationfixed2}
 \{W^{(m)}_i\}_{m\leq n}:= \,\,\{W_i^{(0)},W_i^{(1)},W_i^{(2)},\cdots,W_i^{(n)}\}\,.
\end{gathered}
\end{equation}

We have not defined the probability setting which the cavity field functional is defined on. Let us consider the space $\Omega_r:=\mathbb{R}^{r+1}$ as the sample space of the random variables $\{W^{(m)}\}_{m\leq r}$, endowed with the Borel $\sigma -$algebra $\mathcal{B}_r$ and with the filtration $\{\mathcal{B}^{W}_n\}_{n\leq r}$ such as, for each $0\leq n\leq r$, $\mathcal{B}^{W}_n$ is the $\sigma-$algebra generated by the random variables $\{W^{(m)}\}_{m\leq n}$ (natural filtration)\cite{Billingsley}. Let $\nu$ denote the one dimensional normal distribution and $\mathbb{W}_r:=\nu^{\otimes r}$ be the product probability measure of $r$ normal distribution on $(\Omega_r,\mathcal{B}_r)$. In this formalism, the cavity field functional is a real-valued $\mathcal{B}_r-$measurable function $h:\Omega_r\to \mathbb{R}$.

 We also define the probability spaces $(\Omega_r,\mathcal{B}_r)^{\otimes 2}$ and $(\Omega_r,\mathcal{B}_r)^{\otimes c}$, given, respectively, by the $2-$fold and $c-$fold Cartesian product of the probability space $(\Omega_r,\mathcal{B}_r)$, together with the respectively filtrations $\{(\mathcal{B}^{W}_n)^{\otimes 2}\}_{n\leq r} $ and $\{(\mathcal{B}^{W}_n)^{\otimes c}\}_{n\leq r}$.
 
\subsection{The free energy as a composition of non-linear expectation values}
One of the most remarkable properties of the Ruelle random probability cascade point processes is the quasi-stationarity under a class of time evolutions, which includes the cavity dynamics. In particular, it implies that the average of a population of hierarchical random variables, weighted by the Ruelle random probability cascade configurations $\{\xi_{\bm{\alpha},r}\}_{\bm{\alpha}\in\mathbb{N}^r}$, can be represented with a recursive composition of non-linear expectation values that run only over such variables, getting rid of the cumbersome random weights; this is actually the case of vertex and the edge contribution of the free energy \eqref{cavityFE}.

Note that the edge and the vertex free energy contributions have a quite similar form, then, for simplicity, we will use a unique notation representing both the cases. 

In the rest of this paper, the edge/vertex superscript $\,\cdot\,^{\text{(e/v)}}$ will denote that a given result must be considered both for the two contributions. The symbol $(2/c)$ will denote that one has to consider $2$ or $c$ variables respectively for the edge and vertex contributions. Here, the boldface symbols $\bm{J}$, $\bm{h}_{\bm{\alpha},r}$ and $\bm{W}^{(m)}$represent the arrays of $2$ or $c$ independent random variables, one for each site which the considered function depends on, according to the definitions \eqref{deltaterms}. The quantities without the edge/vertex superscript have the same probability law in both the contributions.

The vertex and the edge contributions verify:
\begin{equation}
\mathbb{E}\log\left(\frac{\sum_{\bm{\alpha}}\xi_{\bm{\alpha}}\Delta^{\text{(e/v)}}(\bm{J},\bm{h}_{\bm{\alpha},r})}{\sum_{\bm{\alpha}}\xi_{\bm{\alpha}}}\right)= \int_{\mathbb{R}^{(2/c)}}\left(\prod^{(2/c)}_{i=1} d\nu\big(W^{(0)}_{i}\big)\,\right) \, \phi^{\text{(e/v)}}_0\big(\bm{J},\bm{W}^{(0)}\big)\, ,
\end{equation}
where $\phi^{\text{(e/v)}}_0\big(\bm{J},\bm{W}^{(0)}\big)$ is the end point of the following backward recursive map:
\begin{equation}
\label{start}
\phi^{\text{(e/v)}}_{r}\big(\,\bm{J},\big\{\bm{W}^{(m)}\big\}_{m\leq r}\,\big)=\log\left(\,\Delta^{\text{(e/v)}}\big(\,\bm{J},\bm{h}_r\big(\,\big\{\bm{W}^{(m)}\big\}_{m\leq r}\big)\,\big)\,\right)\,,
\end{equation}
\begin{multline}
\label{start1}
\phi^{\text{(e/v)}}_{n}\big(\,\bm{J},\big\{\bm{W}^{(m)}\big\}_{m\leq n}\,\big)=\frac{1}{x_{n+1}}\log \mathbb{E}_{\mathbb{W}_r^{\otimes (2/c)}}\left[\,\exp\left(\,x_{n+1} \phi^{\text{(e/v)}}_{n+1}\big(\,\bm{J},\big\{\bm{W}^{(m)}\big\}_{m\leq n+1}\,\big)\,\right)\,\Big|\big\{\bm{W}^{(m)}\big\}_{m\leq n}\right]\\\text{for} 1\leq n\leq r-1
\end{multline}
and finally
\begin{equation}
\label{start2}
\phi^{\text{(e/v)}}_{0}(\bm{J},\bm{W}^{(0)}\,)=\frac{1}{x_{1}}\log \mathbb{E}_{\mathbb{W}_r^{\otimes (2/c)}}\left[\,\exp\left(\,x_{1} \phi^{\text{(e/v)}}_{1}\big(\bm{J},\bm{W}^{(0)},\,\bm{W}^{(1)}\,\big)\,\right)\,\Big|\bm{W}^{(0)}\right]\,,
\end{equation}
where
\begin{equation}
\bm{h}_r\big(\,\big\{\bm{W}^{(m)}\big\}_{m\leq r}\big)=\left(\,h_1\big(\,\big\{W_1^{(m)}\big\}_{m\leq r}\big),\,\cdots\,,h_{c}\big(\,\big\{W_c^{(m)}\big\}_{m\leq r}\big)\,\right)\,,
\end{equation}
and
\begin{equation}
\bm{h}_r\big(\,\big\{\bm{W}^{(m)}\big\}_{m\leq r}\big)=\left(\,h_1\big(\,\big\{W_1^{(m)}\big\}_{m\leq r}\big),\,h_{2}\big(\,\big\{W_2^{(m)}\big\}_{m\leq r}\big)\,\right)\,,
\end{equation}
respectively for the vertex and adge contribution.

The symbol $\mathbb{E}_{\mathbb{W}_r^{\otimes (2/c)}}[\cdot|\{\bm{W}^{(m)}\}_{m\leq n}]$ is the expectation over the variables $\bm{W}^{(n+1)},\bm{W}^{(n+2)},\cdots,\bm{W}^{(r)}$, taking the realization of the random variables $\bm{W}^{(0)},\bm{W}^{(1)},\cdots,\bm{W}^{(n)}$ fixed. The symbol $\mathbb{W}_r^{\otimes (2/c)}$ denotes the probability measure given by the $2-$fold or $c-$fold (according to the case) product of the probability measure $\mathbb{W}_r$. 
The functional $\phi^{\text{(e/v)}}_{n}$, for each level $0\leq n\leq r$, depends only on the first $n$ random variables $\{\bm{W}^{(m)}\}_{m\leq n}$, so we say that the process $\phi^{\text{(e/v)}}:=\{\phi^{\text{(e/v)}}_{n}(\bm{J},\{\bm{W}^{(m)}\}_{m\leq n}\,)\}_{1\leq n\leq r}$ is adapted (or non-anticipative) to the filtration of product $\sigma-$algebras $\{(\mathcal{B}^{W}_n)^{\otimes (2/c)}\}_{n\leq r}$; in particular it is a supermartingale \cite{YoRev}. For each level n, with $0\leq n\leq r$, we refers to the the first $n$ random variables $\{\bm{W}^{(m)}\}_{m\leq n}$ as \emph{past random variables}.

Let us define also the $r-$steps free energy stochastic process $\phi$ such that:
\begin{equation}
\phi=\phi^{\text{(v)}}-\frac{c}{2}\phi^{\text{(e)}}
\end{equation}

Note that the process $\phi$ depends on the variables $\{\bm{W}^{(m)}\}_{m\leq r}$ through the cavity field functional $h_r$.

In the Parisi solution of the SK model \cite{VPM}, the cavity field functional is linear, so the free energy process is Markovian. In this case, at each level $0\leq n\leq r$, the functional $\phi_n$ depends on the random variables $\{\bm{W}^{(m)}\}_{m\leq n}$ only through the linear combination
\begin{equation}
h_n=\sum^n_{m=1}\sqrt{q_{m}-q_{m-1}} \,\bm{W}^{(m)}\,,
\end{equation}
where $q_1,q_2,\cdots, q_r$ are the overlaps \cite{VPM,ASS}. 

The functional $\phi_n$ is actually a function of one variable, for all the levels $0\leq n\leq r$ and for any number of RSB steps $r\in\mathbb{N}$. The expectation $\mathbb{E}_{\mathbb{W}_r^{\otimes (2/c)}}[\cdot|\{\bm{W}^{(m)}\}_{m\leq n}]$ in \eqref{start1} can be substituted by the expectation over $h_{n+1}$, conditionally to $h_n$. As a consequence, for each level $0\leq n\leq r$, the expectation value of $\phi_{n+1}$ can be evaluated through the Kolmogorov backward equation, which is a deterministic (non-random) partial differential equation (PDE) \cite{YoRev}. After some manipulation, the function $\phi_n$ can be write as the solution at time $q_{n}$ of a proper PDE, starting from $\phi_{n+1}$ at time $q_{n+1}$. The juxtaposition of such PDEs provides a “continuous version" of the iteration \eqref{start1} and defines a backward map $\phi_r\mapsto \phi_0$, that leads to the Parisi antiparabolic PDE in the $r\to \infty$ limit\cite{VPM}.

As suggested by Parisi in \cite{ParisiMarginal}, a similar construction can be generalized to a wider class of cavity field functional $h_r:\Omega_r\to \mathbb{R}$, provided that, at least in the $r\to\infty$ limit, the process $H:=\{H_n(\{W^{(m)}\}_{m\leq n})\}_{l\leq r}$, defined as
\begin{equation} 
H_r(\{W^{(m)}\}_{m\leq n})=h_r(\{W^{(m)}\}_{m\leq n})
\end{equation}
and
\begin{equation} 
H_n(\{W^{(m)}\}_{m\leq n})=\mathbb{E}_{\mathbb{W}_r^{\otimes (2/c)}}[h_r(\{W^{(m)}\}_{m\leq n})\,|\{W^{(m)}\}_{m\leq n}]\,,
\end{equation}
is a Markov martingale. In this case, a Parisi-like equation can be achieved from the master equation of the process $H$ \cite{YoRev}.

For more generic cavity field functional, the free energy process is not Markovian, i.e. , for each level $n$, the functional $\phi_n$ has a detailed dependence on the specific values of each past variable of the list $\{\bm{W}^{(m)}\}_{m\leq n}$. Non-Markovianity is the basic difference with the replica symmetry breaking scheme in the fully connected model \cite{VPM,ParisiMarginal}. 

Because non-Markovianity, we cannot get rid of the randomness represented by the past variables, so the free energy process cannot be the solution of a deterministic partial differential equation. Furthermore, in the $r\to \infty$ limit, the free energy process depends on an infinite number of variables. One may consider a functional extension of the Parisi PDE, based on the functional It\^o calculus and functional Kolmogorov equations \cite{Dupire,ContFur}. However, any continuous version of the iteration \eqref{start1} would depend on the cavity field functional in a tricky way, then the dependence of the functional $\phi_0$ on the cavity field functional would be not explicit, so the variational problem on such parameter turns out to be quite cumbersome with this approach.

The approach described in the next section evaluates directly the map $\phi_r\mapsto \phi_0$, avoiding any iteration procedure as in \eqref{start1}.
 
In the following, the dependence of the process $\phi^{\text{(e)}}$ and $\phi^{\text{(v)}}$ on the random couplings will be omitted for convenience; all the equations below will refer to a single realization of the random couplings.
\subsection{Variational representation of the $r$-RSB free energy functional}
In this subsection, we get a variational representation of the recursive law \eqref{start}. The variational representation turns out to be a powerful tool to get the $r\to \infty$ limit. 

In this subsection, and in the rest of the paper, the martingale formalism \cite{YoRev} is deeply used.
Let first introduce some notation.

Let $\mathfrak{D}^{\text{(e/v)}}_r$ be the set of the stochastic processes adapted to the filtration $\{(\mathcal{B}^{W}_n)^{\otimes (2/c)}\}_{n\leq r}$ (both for edge and vertex contribution).

Let $\mathfrak{M}^{\text{(e/v)}}_{r}\subset \mathfrak{D}^{\text{(e/v)}}_r$ the subspace of $\{(\mathcal{B}^{W}_n)^{\otimes (2/c)}\}_{n\leq r}-$adapted martingales and $\mathfrak{M}^{\text{(e/v)}}_{r,1,>}\subset \mathfrak{M}^{\text{(e/v)}}_r$ the subset of strictly positive $\{(\mathcal{B}^{W}_n)^{\otimes (2/c)}\}_{n\leq r}-$adapted martingales with average equal to $1$.

Furthermore, for each level $0<n\leq r$, let $\mathfrak{R}^{\text{(e/v)}}_{n,1,>}$ be the set of strictly positive random variables, depending on the random variables $\big\{\bm{W}^{(m)}\big\}_{m\leq n}$, with expectation value over $\bm{W}^{(n)}$, conditionally to a fixed realization of the variables $\big\{\bm{W}^{(m)}\big\}_{m\leq n-1}$, equal to $1$.

The second member of the recursion formula \eqref{start} has the form of the usual Helmotz free energy in the canonical ensemble, with $x_{n+1}$ as inverse temperature and $-\phi_{n+1}$ as Hamiltonian. As a consequence, it can be represented via the Gibbs variational principle.

For each level $n\leq r$ and each fixed realization of the first $n$ random variables $\big\{\bm{W}^{(m)}\big\}_{m\leq n}$, let us generate a strictly positive random variable $\rho_{n+1}^{\text{(e/v)}}(\,\Cdot\,|\,\{\bm{W}^{(m)}\}_{m\leq n}\,\big)$, depending on the random variable $\bm{W}^{(n+1)}$, satisfying the following normalization condition:
\begin{equation}
\label{normalRho}
\mathbb{E}_{\mathbb{W}_r^{\otimes (2/c)}}\left[\rho_{n+1}^{\text{(e/v)}}\big(\bm{W}^{(n+1)}|\,\{\bm{W}^{(m)}\}_{m\leq n}\,\big) \,\Big|\{\bm{W}^{(m)}\}_{m\leq n}\right]=1\,\, \longrightarrow\,\, \rho_{n+1}^{\text{(e/v)}}\in \mathfrak{R}^{\text{(e/v)}}_{n,1,>}\,.
\end{equation}
The variable $\rho_{n+1}^{\text{(e/v)}}$, actually, plays the role of an effective conditional probability density function for the variable $\bm{W}^{(n+1)}$, given the realization of the past variables $\big\{\bm{W}^{(m)}\big\}_{m\leq n}$. The Gibbs principle provides a variational criterion on the space of the density functions
\begin{multline}
\label{iterVariatinal}
\phi^{\text{(e/v)}}_{n}\big(\{\bm{W}^{(m)}\}_{m\leq n}\,\big)=\\\max_{\rho_{n+1}^{\text{(e/v)}}\in\mathfrak{R}^{\text{(e/v)}}_{n,1,>}}\bigg\{
\mathbb{E}_{\mathbb{W}_r^{\otimes (2/c)}}\left[\rho_{n+1}^{\text{(e/v)}}\big(\bm{W}^{(n+1)}|\,\{\bm{W}^{(m)}\}_{m\leq n}\,\big) \phi_{n+1} ^{\text{(e/v)}}\big(\{\bm{W}^{(m)}\}_{m\leq n+1}\,\big)\,\Big|\{\bm{W}^{(m)}\}_{m\leq n}\right]\\
-\frac{1}{x_{n+1}}\mathbb{E}_{\mathbb{W}_r^{\otimes (2/c)}}\left[\rho_{n+1}^{\text{(e/v)}}\big(\bm{W}^{(n+1)}|\,\{\bm{W}^{(m)}\}_{m\leq n}\,\big)\log\,\rho_{n+1}^{\text{(e/v)}}\big(\bm{W}^{(n+1)}|\,\{\bm{W}^{(m)}\}_{m\leq n}\,\big)\,\Big|\{\bm{W}^{(m)}\}_{m\leq n}\right] \bigg\}\,,
\end{multline}
where the maximum is attained by:
\begin{multline}
\label{extremal}
\rho_{n+1}^{(e/v)\star}\big(\bm{W}^{(n+1)}|\,\{\bm{W}^{(m)}\}_{m\leq n}\,\big)=\frac{1}{Z^{\text{(e/v)}}_{n}\big(\{\bm{W}^{(m)}\}_{m\leq n}\,\big)}\exp\left(x_{n+1}\phi_{n+1} ^{\text{(e/v)}}\big(\{\bm{W}^{(m)}\}_{m\leq n+1}\,\big)\,\right),\\
\text{with}\quad Z_{n}\big(\{\bm{W}^{(m)}\}_{m\leq n}\,\big)=\mathbb{E}_{\mathbb{W}_r^{\otimes (2/c)}}\left[\exp\left(x_{n+1}\phi_{n+1} ^{\text{(e/v)}}\big(\{\bm{W}^{(m)}\}_{m\leq n+1}\,\big)\,\right)\Big|\{\bm{W}^{(m)}\}_{m\leq n}\right]\,.
\end{multline}

The non-linear maps $\phi_{n} ^{\text{(e/v)}}\,\mapsto \phi_{n-1} ^{\text{(e/v)}}$ in \eqref{start} are now represented as linear maps \eqref{extremal}, with the help of suitable variational parameters. 

In physics literature, the first part in the representation \eqref{iterVariatinal}, depending on $\phi_{n+1}^{e/v}$, is called ``energy", whilst the second one, with the logarithm, is the ``entropic" part.

Such kind of manipulation is also at the basis of the Boué-Dupuis representation formula for the expectation value of exponential Brownian functionals\cite{BoueDepuis}.

From the effective conditional density functions $\{\,\rho_{1}^{\text{(e/v)}}, \rho_{2}^{\text{(e/v)}},\cdots,\rho_{r}^{\text{(e/v)}}\}\in \mathfrak{R}^{\text{(e/v)}}_{0,1,>}\times \mathfrak{R}^{\text{(e/v)}}_{1,1,>}\times \cdots \times \mathfrak{R}^{\text{(e/v)}}_{r,1,>}$, we can compute an effective probability density function $R_r^{\text{(e/v)}}$ for the whole collection of random varables $\{\bm{W}^{(n)}\}_{n\leq r}$:
\begin{equation}
R_r^{\text{(e/v)}}\big(\{\bm{W}^{(m)}\}_{m\leq r}\big)=\prod_{m=1}^r \rho_m^{\text{(e/v)}} \big(\bm{W}^{(m)}|\{\bm{W}^{(l)}\}_{l\leq m-1}\big)\,.\\
\end{equation}
By condition \eqref{normalRho}, the function $R_r^{\text{(e/v)}}$ is normalized:
\begin{equation}
\label{RtoRho}
\mathbb{E}_{\mathbb{W}_r^{\otimes (2/c)}}\left[R_r^{\text{(e/v)}}\big(\{\bm{W}^{(m)}\}_{m\leq r}\big)\right]=1.\\
\end{equation}

For any level $0\leq n \leq r$, the effective marginal probability density $R_n^{\text{(e/v)}}$ over the first $n$ random variables $\{\bm{W}^{(m)}\}_{m\leq n}$ is given by averaging $R_r^{\text{(e/v)}}$ over the last $r-n$ random variables $\{\bm{W}^{(m)}\}_{n\leq m\leq r}$
\begin{equation}
\label{marginalR}
R_n^{\text{(e/v)}}\big(\{\bm{W}^{(m)}\}_{m\leq n}\big)=\mathbb{E}_{\mathbb{W}_r^{\otimes (2/c)}}\left[R_r^{\text{(e/v)}}\big(\{\bm{W}^{(m)}\}_{m\leq r}\big) \Big|\{\bm{W}^{(m)}\}_{m\leq n}\,\right]=\prod_{m=1}^n \rho_m^{\text{(e/v)}} \big(\bm{W}^{(m)}|\{\bm{W}^{(l)}\}_{l\leq m-1}\big)
\end{equation}
and
\begin{equation}
\label{R0}
R_0^{\text{(e/v)}}\big(\bm{W}^{(0)}\big)=1
\end{equation}

All the marginal density functions, defined by \eqref{marginalR} and \eqref{R0}, are already normalized, by construction.

The ``entropic" part in \eqref{iterVariatinal} can be rewitten as a functional of the probability density function $R_r^{\text{(e/v)}}$ and the marginals, in such a way:
\begin{equation}
\begin{aligned}
&\mathbb{E}_{\mathbb{W}_r^{\otimes (2/c)}}\left[\rho_{n+1}^{\text{(e/v)}}\big(\{\bm{W}^{(m)}\}_{m\leq n+1}\,\big)\log\,\rho_{n+1}^{\text{(e/v)}}\big(\{\bm{W}^{(m)}\}_{m\leq n+1}\,\big)\,\Big|\{\bm{W}^{(m)}\}_{m\leq n}\right]=\\
& \! \begin{multlined}
\frac{1}{R_n^{\text{(e/v)}}\big(\{\bm{W}^{(m)}\}_{m\leq n}\big)}\mathbb{E}_{\mathbb{W}_r^{\otimes (2/c)}}\bigg[\,R_{n+1}^{\text{(e/v)}}\big(\{\bm{W}^{(m)}\}_{m\leq n+1}\big)\\ \times\bigg(\,\log\Big(R_{n+1}^{\text{(e/v)}}\big(\{\bm{W}^{(m)}\}_{m\leq n+1}\big)\,\Big)-\log\Big(R_{n}^{\text{(e/v)}}\big(\{\bm{W}^{(m)}\}_{m\leq n}\big)\,\Big)\,\bigg)\,\bigg| \{\bm{W}^{(m)}\}_{m\leq n}\bigg]=
\end{multlined}\\
& \! \begin{multlined}
\frac{1}{R_n^{\text{(e/v)}}\big(\{\bm{W}^{(m)}\}_{m\leq n}\big)}\mathbb{E}_{\mathbb{W}_r^{\otimes (2/c)}}\bigg[\,R_{r}^{\text{(e/v)}}\big(\{\bm{W}^{(m)}\}_{m\leq r}\big)\\ \times\bigg(\,\log\Big(R_{n+1}^{\text{(e/v)}}\big(\{\bm{W}^{(m)}\}_{m\leq n+1}\big)\,\Big)-\log\Big(R_{n}^{\text{(e/v)}}\big(\{\bm{W}^{(m)}\}_{m\leq n}\big)\,\Big)\,\bigg)\,\bigg| \{\bm{W}^{(m)}\}_{m\leq n}\bigg]
\end{multlined}
\end{aligned}
\end{equation}

The monotonicity of the expectation value on $\mathfrak{D}^{\text{(e/v)}}_r$ and the tower property \cite{YoRev}\cite{Billingsley}, the iteration of the representations \eqref{iterVariatinal} leads to a unique variational representation for the entire map $\phi_{r} ^{\text{(e/v)}}\,\mapsto \phi_{0} ^{\text{(e/v)}}$:

\begin{multline}
\label{GibbsFunctional}
\phi^{\text{(e/v)}}_{0}(\bm{W}^{(0)}\,)=
\max_{R^{\text{(e/v)}}\in \mathfrak{M}^{\text{(e/v)}}_{r,1,>}}
\Bigg\{\mathbb{E}_{\mathbb{W}_r^{\otimes (2/c)}}\left[R_r^{\text{(e/v)}}\big(\{\bm{W}^{(m)}\}_{m\leq r}\big) \phi_{r} ^{\text{(e/v)}}\big(\{\bm{W}^{(m)}\}_{m\leq r}\big)\,\Big| \bm{W}^{(0)}\, \right]\\-\mathbb{E}_{\mathbb{W}_r^{\otimes (2/c)}}\left[\,R_r^{\text{(e/v)}}\big(\{\bm{W}^{(m)}\}_{m\leq r}\big)\sum_{n=1}^{r}\frac{1}{x_n}\bigg(\,\log\Big(R_n^{\text{(e/v)}}\big(\{\bm{W}^{(m)}\}_{m\leq n}\big)\,\Big)-\log\Big(R_{n-1}^{\text{(e/v)}}\big(\{\bm{W}^{(m)}\}_{m\leq n-1}\big)\,\Big)\,\bigg)\,\bigg| \,\bm{W}^{(0)}\,\right]\,\Bigg\}\,.
\end{multline}

The maximum is formally attained substituting the solutions \eqref{extremal} in \eqref{RtoRho}, for each level $n$.

Note that the collection of effective marginal probability densities, defined in \eqref{marginalR}, defines an average $1$ strictly positive martingale $R^{\text{(e/v)}}:=\big\{R_n^{\text{(e/v)}}\big(\{\bm{W}^{(m)}\}_{m\leq n}\big)\big\}_{n\leq r}\in \mathfrak{M}^{\text{(e/v)}}_{r,1,>}$. The martingale property, indeed, is stated by the definition of marginal density functions in the equation \eqref{marginalR}.

The Gibbs variational principle, then, combines a cumbersome recursive composition of conditional non-linear expectations values in a single variational problem over the space $\mathfrak{M}^{\text{(e/v)}}_{r,1,>}$ of positive, average $1$, martingales on the filterd probability space $(\Omega_r,\mathcal{B}_r,\{(\mathcal{B}^{W}_n)\}_{n\leq r},\mathbb{W}_r)^{\otimes (2/c)}$. This is a big deal of such approach, since martingales are well-defined mathematical object in any generic probability space. 

The total free energy functional is given by the sum of the edge and vertex contributions, averaged over the root random variables $\bm{W}^{(0)}$ and the random couplings:
\begin{equation}
\label{GibbsFunctional2}
\Phi= \int_{\mathbb{R}^{c}}\left(\prod^{c}_{i=1} d\nu\big(W^{(0)}_{i}\big)\,\right) \, \phi^{\text{(v)}}_0\big(\bm{J},\bm{W}^{(0)}\big)-\frac{c}{2}\int_{\mathbb{R}^{2}}\left(\prod^{2}_{i=1} d\nu\big(W^{(0)}_{i}\big)\,\right) \, \phi^{\text{(e)}}_0\big(\bm{J},\bm{W}^{(0)}\big)\, .
\end{equation}

The equilibrium free-energy is given by the extremization of the total free energy functional \eqref{GibbsFunctional2} with respect to the physical order parameter, i.e. the cavity field functional $h_r\in \mathfrak{F}(\Omega_r,\mathbb{R})$; the representations of the two free energy contributions, given by\eqref{GibbsFunctional}, constitute two independent variational problems inside a larger variational problem.

In the following, we will refer to \eqref{GibbsFunctional} as edge and vertex auxiliary variational problems, whilst the extremization over the cavity field functional is the physical variational problem. The auxiliary variational problems must be solved before, keeping the variational parameters of the physical variational problem fixed.

The careful reader may argue that the Gibbs variational principle seems not to provides a real simplification since the computation of the non-linear expectations still remains in the solutions \eqref{extremal}. This is actually true in the discrete replica symmetry breaking case.

In the ``continuous limit" of replica symmetry breaking, however, the martingale $R^{\text{(e/v)}}$ has a nice representation and we do not need to deal with the solution \eqref{extremal}, as explained in the next section.

\section{The continuous replica symmetry breaking}
In the previous section, the $r$-RSB free energy \eqref{cavityFE} is represented as a variational problem over the space of positive discrete time martingales.

In this section, the auxiliary variational problem \eqref{GibbsFunctional} is extended to continuous-time martingales that have also a continuous sample path \cite{YoRev} and get the full-RSB free energy functional.

 From a rigorous mathematical point of view, such extension should not be considered as the $r\to \infty$ limit of the $r$-RSB theory, but rather a generalization of the previous stochastic analysis to another class of martingales.

In the first subsection, we present the formal derivation of the auxiliary variational problem in the continuous case.

In the second subsection, we reduce the analysis to It\^o stochastic processes, for which the auxiliary variational problem is solvable, and the full-RSB free energy functional is finally achieved.

We then address the physical variational problem: we define a suitable order parameter and derive the mean-field equations from the stationary condition of the free energy functional.

\subsection{The generalized Chen-Auffinger variational representation}
In this subsection, the variational auxiliary problem for continuous martingales is derived.

At the first, we rephrase the martingale approach of the previous section, with a suitable formalism, representing both continuous and discrete martingales. Then we concentrate on path continuous martingales, getting a generalization of Auffinger-Chen variational representation of the Parisi solution of the SK models \cite{ChenAuf}.

\paragraph{continuous-time formalism.} Given any ordered collection of random variables $\{W^{(m)}\}_{m\leq r} \in \Omega_r$, together with the increasing sequence $X=(0=x_0\leq x_1\leq \cdots \leq x_{r+1}=1)$, we can define a continuous-time, stepwise, bounded random function $W_{\text{c}}:=\{W_{\text{c}}(x);\,x\in [0,1]\}$, such as for each level $0\leq n\leq r$ and time $x_{n}< x\leq x_{n+1}$, the random quantity $W_{\text{c}}(x)$ depends only on the first $n$ variables $\{W^{(m)}\}_{m\leq n}$. A possible choice may be
\begin{equation}
\label{continuation}
W_{\text{c}}(x)= W^{(0)}1_{\{0\}}(x)+\sum_{i=1}^{r} \sqrt{x_{n}-x_{n-1}}\, \,W^{(n)} \theta(x-x_{n-1}),\quad x\in [0,1],
\end{equation}
 where the function $\theta$ is the Heaviside function and the function $1_{\{0\}}(x)$ has the value $1$ at $x=0$ and the value $0$ for all $x>0$.

Any $r-$ step adapted process $O:=\{O_0,O_1,\cdots,O_r\} \in \mathfrak{D}^{(e/v)}_r$ can now be considered as an ordered collection of functionals of the vector random function $\bm{W_{\text{c}}}$, constituted by $2$ or $c$ independent realizations of the random function defined in \eqref{continuation}:
\begin{equation}
O_n=O_n\big(\,\{\bm{W}^{(m)}\}_{m\leq n}\,\big)\longrightarrow O_n\big(\{\bm{W}_\text{c}(x);0\leq x\leq x_n\}\big)
\end{equation}
where the symbol $\{\bm{W}_\text{c}(x);0\leq x\leq x_n\}$ denotes that the functional $O_n$ depends on all the values $\bm{W}_\text{c}(x)$, assumed by the realization of the random function $\bm{W}_\text{c}$ at each time $x\in[0,x_n]$.

We define a continuous-time process, depending on $\bm{W}_{\text{c}}$ and based on $O$, in such a way:
\begin{equation}
\label{continuation2}
O_{\text{c}}(x)= O_0 1_{[0,x_1]}(x)+\sum_{i=1}^{r} O_n 1_{(x_{n},x_{n+1}]}(x),\quad x\in [0,1].
\end{equation}
where the functions $1_{(x_{n},x_{n+1}]}(x)$, for $1\leq n\leq r$, are equal to $1$ if $x_{n-1}<x\leq x_{n}$ and vanish elsewhere.

With this construction of the continuous-time processes, the values of $O_{\text{c}}(x)$, at each time $0\leq x\leq 1$, depends only on the values $\bm{W}_{\text{c}}(x')$, assumed by a given realization of the stepwise random function $\bm{W}_{\text{c}}$, at all the times $x'\in[0,x]$. So the process $O_{\text{c}}$ is adapted to the continuous-time stepwise random function $\bm{W}_{\text{c}}$. 

In the following, the subscript $\Cdot_{\text{c}}$ will be omitted and we will deal only with continuous-time stochastic process.

In the $r\to\infty$ limit, the sequence $X$ ``fills" the $[0,1]$ segment, so this limit can be formally achieved enlarging the space of stepwise processes to a wider class of continuous-time processes.

The probability space will be carefully defined in the last paragraph of this section. For now, let $\Omega$ be a generic sample space of functions for the process $W$, with a proper $\sigma-$algebra $\mathcal{B}$ and a given probability measure $\mathbb{W}$.

A process $O:\Omega^{\otimes(2/c)}\to \mathbb{R}$ is said to be adapted (or non-anticipating) if it is adapted with respect to the usual $\mathbb{W}^{\otimes (2/c)}$-completion \footnote{The usual completion of a continuous-time filtration, with respect to a given probability measure, is the smallest right-continuous filtration that contains the original one, enlarged with the set with probability $0$ with respect to the given probability measure. Usually, a rigorous treatment of continuous-time stochastic processes requires the usual completion.}of the natural filtration of the vector process $\bm{W}$ on the probability space $(\Omega,\mathcal{B},\mathbb{W})^{\otimes(2/c)}$.

Let us remind some standard crucial mathematical tools, defined for continuous-time stochastic processes, that will be useful in the next paragraphs \cite{YoRev}. 

We call quadratic variation of a process $O$ the process $[O]$ defined as:
\begin{equation}
\label{var}
[O](x)=\lim_{\substack{M\to \infty\\ \delta \to 0}} \sum^M_{n=0} \left(O(x_{n+1})-O(x_{n})\right)^2,\,\,\text{with} \,\,0=x_0\leq x_1\leq \cdots\leq x_M=x\,,
\end{equation}
where $\delta$ is the mesh of the partition. 

For stepwise processes, as \eqref{continuation2}, this quantity is reduced to the sum over the discontinuity jumps in such a way:
\begin{equation}
\label{varfinit}
[O](x)= \sum^r_{n=1} 1_{[x_{n},x_{n+1})}(x)\left(O_{n}-O_{n-1}\right)^2,\,\,\text{with}\, 0=x_0\leq x_1\leq \cdots\leq x_r=1\,.
\end{equation}

In the same manner, the covariation between two processes $O$ and $P$ is the process $\Braket{O,P}$ such that:
\begin{equation}
\label{covar}
\Braket{O,P}(x)=\lim_{\substack{M\to \infty\\ \delta \to 0}} \sum^M_{n=0} \left(O(x_{n+1})-O(x_{n})\right)\left(P(x_{n+1})-P(x_{n})\right),\,\,\text{with}\, 0=x_0\leq x_1\leq \cdots\leq x_M=x\,.
\end{equation}
Obviously, the quadratic variation and the covariation vanish for smooth functions, but they are not trivial for all continuous martingale.

We can also define the integration with respect to a continuous-time stochastic process $O$ by the Lebesgue-Stieltjes integral \cite{Billingsley} or by the It\^o integral \cite{Ito} whether the integration process $O$ is of bounded variation or of bounded quadratic variation respectively\cite{YoRev}.

Note that, by definition, for any bounded function $f:[0,1]\to \mathbb{R}$ and stepwise process $O$, we have:
\begin{equation}
\int^1_0 f(x) dO(x)=\sum^r_{n=1} f(x_{n-1})(\,O_{n}-O_{n-1}\,).
\end{equation}

Thank to this formalism, the auxiliary variational representation \eqref{GibbsFunctional} can be easily reformulated in term of generic continuous-time processes as:
\begin{multline}
\label{GibbsFunctional3}
\phi^{\text{(e/v)}}(0,\bm{W}{(0)}\,)=
\max_{R^{\text{(e/v)}}\in \mathfrak{M}^{\text{(e/v)}}_{[0,1],1,>}}
\Bigg\{\mathbb{E}_{\mathbb{W}^{\otimes(2/c)}}\left[ R^{\text{(e/v)}}\left(\,1,\,\bm{W}\,\right)\phi ^{\text{(e/v)}}\left(\,1,\,\bm{W}\,\right)\,\bigg| \,\bm{W}{(0)} \right]\\-\mathbb{E}_{\mathbb{W}^{\otimes (2/c) }}\left[R^{\text{(e/v)}}\left(\,1,\,\bm{W}\,\right)\int^1_0 \frac{1}{x}\,d\left(\,\log\, R^{\text{(e/v)}}\left(\,x,\,\bm{W}\,\right)\,\,\right)\,\,\bigg| \,\bm{W}{(0)}\,\right]\,\Bigg\}\,,
\end{multline}
where 
\begin{equation}
\phi ^{\text{(e/v)}}\left(\,1,\,\bm{W}\,\right)=\log \,\Delta^{\text{(e/v)}} \left(\,\bm{h}\left(\,\{\bm{W}(x);0\leq x\leq 1\}\,\right)\,\right)
\end{equation}
and we have used the short-hand notation
\begin{equation}
\label{short}
R^{\text{(e/v)}}\left(\,x,\,\bm{W}\,\right)=R^{\text{(e/v)}}\left(\,x,\,\{\bm{W}(x);0\leq x'\leq x\}\,\right)\,,\quad \forall 0 \leq x \leq 1.
\end{equation}
The range set $\mathfrak{M}^{\text{(e/v)}}_{[0,1],1,>}$ of the variational parameter $R^{\text{(e/v)}}$ is the space of positive martingales on the probability space $(\Omega,\mathcal{B},\mathbb{W})^{\otimes(2/c)}$, with average $1$. 

The representation \eqref{GibbsFunctional3} holds both for stepwise and continuous martingales, providing a general formulation for all levels of discrete replica symmetry breaking and for the full replica symmetry breaking.

Note that the auxiliary variational representation turns out to be a powerful tool in getting the continuous limit of replica symmetry breaking free energy functional. The extension to the continuous case of the representation \eqref{GibbsFunctional2} has been easily defined in \eqref{GibbsFunctional3}, simply by introducing a proper notation, allowing us to deal with quantities that are well defined even in the $r\to \infty$ limit. By contrast, the $r\to \infty$ limit of the iterative low \eqref{start1} appears to be quite cumbersome, since the cavity field functional \eqref{AustinRep} depends on the random quantities $\{W^{(m)}\}_{m\leq r}$ in a non trivial way.

The set $\mathfrak{M}^{\text{(e/v)}}_{[0,1],1,>}$ is too generic for many practical computation, so in the next paragrahs further assumptions on the martingale $R^{\text{(e/v)}}$ will be imposed.

\paragraph{Path continuity assumption.} A generic martingale $R^{\text{(e/v)}}\in\mathfrak{M}^{\text{(e/v)}}_{[0,1],1,>}$ can be decomposed as the sum of a purely continuous martingale and a purely discontinuous martingale.

In the $r-$step replica symmetry breaking, the continuous part vanishes, and the martingale $R^{\text{(e/v)}}$ is purely discontinuous. Here, we are interested in the case where $R^{\text{(e/v)}}$ is purely continuous. 

For the rest of the paragraph, we will use the shorthand notation$(x,\bm{W})$ to indicate the non-anticipating dependence of the processes to the random function $\bm{W}$, as in \eqref{short}.

Any strictly positive and path continuous martingale, can be represented as the exponential function of a path continuous supermartingale \cite{YoRev}. More precisely, using the It\^o lemma of the stochastic differential calculus \cite{Ito}, the martingale condition and the normalization of $R^{\text{(e/v)}}$ lead to the following representation ( proposition 1.8 in Chapter 8 of \cite{YoRev}):
\begin{equation}
\label{exp}
R^{\text{(e/v)}}\left(\,x,\bm{W}\,\right)=\mathcal{E}_x\left(\,L^{\text{(e/v)}}\,\right)=e^{L^{\text{(e/v)}}(x,\bm{W})-\frac{1}{2}[L^{\text{(e/v)}}](x,\bm{W})}\,,
\end{equation}
where $L^{\text{(e/v)}}(x,\bm{W})$ is a sample continuous martingale with
\begin{equation}
\label{0Cond}
\mathbb{E}_{\mathbb{W}^{\otimes (2/c) }}\left[L^{\text{(e/v)}}(x,\bm{W})\Big| \,\bm{W}^{(0)}\right]=0\,
\end{equation}
and $\mathcal{E}_x$ is the Doléans-Dade exponential \cite{YoRev} and 

Using the stochastic integral representation of the exponential \eqref{exp}
\begin{equation}
\label{stocInt}
R^{\text{(e/v)}}(1,\bm{W})=1+\int^1_0R^{\text{(e/v)}}(1,\bm{W}) \,dL^{\text{(e/v)}}(x,\bm{W})
\end{equation}
and substituting the equations \eqref{0Cond}, \eqref{exp} and \eqref{stocInt} in the entopic part of \eqref{GibbsFunctional3}, one gets
\begin{multline}
\mathbb{E}_{\mathbb{W}^{\otimes (2/c) }}\left[R^{\text{(e/v)}}(1,\bm{W})\int^1_0 \frac{1}{x}\,d\left(\,\log\, R^{\text{(e/v)}}(x,\bm{W})\,\,\right)\,\,\bigg| \,\bm{W}^{(0)}\,\right]=\\
\mathbb{E}_{\mathbb{W}^{\otimes (2/c) }}\left[\,R^{\text{(e/v)}}(1,\bm{W})\int^1_0 \frac{1}{x}\,dL^{\text{(e/v)}}(x,\bm{W})\,\bigg| \,\bm{W}^{(0)}\,\right]-\frac{1}{2}\mathbb{E}_{\mathbb{W}^{\otimes (2/c) }}\left[\,R^{\text{(e/v)}}(1,\bm{W})\int^1_0 \frac{1}{x}\,d[L]^{\text{(e/v)}}(x,\bm{W})\,\,\bigg| \,\bm{W}^{(0)}\,\right]=\\
\mathbb{E}_{\mathbb{W}^{\otimes (2/c) }}\left[\,\int^1_0 \frac{1}{x}\,\,dL^{\text{(e/v)}}(x,\bm{W})\,\,\,\,\bigg| \,\bm{W}^{(0)}\,\right]+\mathbb{E}_{\mathbb{W}^{\otimes (2/c) }}\left[\,\int^1_0 R^{\text{(e/v)}}(x,\bm{W})dL^{\text{(e/v)}}(x,\bm{W})\,\int^1_0 \frac{1}{x}\,\,dL^{\text{(e/v)}}(x,\bm{W})\,\,\,\,\bigg| \,\bm{W}^{(0)}\,\right]\\
-\frac{1}{2}\mathbb{E}_{\mathbb{W}^{\otimes (2/c) }}\left[\int^{1}_0 \frac{1}{x}\,\mathbb{E}_{\mathbb{W}^{\otimes (2/c) }}\left[R^{\text{(e/v)}}(1,\bm{W})\,\Big| \,\{\bm{\omega}\}_{x}\,\right]d[L]^{\text{(e/v)}}(x,\bm{W})\bigg| \,\bm{W}^{(0)}\,\right]=\\
\frac{1}{2}\mathbb{E}_{\mathbb{W}^{\otimes (2/c) }}\left[\int^{1}_0 \frac{1}{x}\,R^{\text{(e/v)}}(x,\bm{W})d[L]^{\text{(e/v)}}(x,\bm{W})\bigg| \,\bm{W}^{(0)}\,\right]\,
\end{multline}
and the functional \eqref{GibbsFunctional3} becomes:
\begin{multline}
\label{GibbsFunctional4}
\phi^{\text{(e/v)}}(0,\bm{W}{(0)}\,)=
\max_{L^{\text{(e/v)}}\in \mathfrak{M}^{\text{(e/v)}}_{[0,1],C}}
\Bigg\{\mathbb{E}_{\bm{h},\mathbb{W}^{\otimes (2/c)}}\left[e^{L^{\text{(e/v)}}(1,\bm{W})-\frac{1}{2}[L^{\text{(e/v)}}](1,\bm{W})}\phi ^{\text{(e/v)}}\left(\,1,\,\bm{W}\,\right)\,\bigg| \,\bm{W}^{(0)} \right]\\-\frac{1}{2}\mathbb{E}_{\bm{h},\mathbb{W}^{\otimes(2/c)}}\left[\int^1_0\frac{1}{x}\,\, e^{L^{\text{(e/v)}}(x,\bm{W})-\frac{1}{2}[L^{\text{(e/v)}}](x,\bm{W})}\,d[\,L^{\text{(e/v)}}\,](x,\bm{W})\,\,\bigg| \,\bm{W}^{(0)}\,\right]\,\Bigg\},
\end{multline}
where $\mathfrak{M}^{\text{(e/v)}}_{[0,1],C}$ is the set of $\bm{h}-$adapted martingales under the probability measure $\mathbb{W}^{\otimes(2/c)}$, with continuous sample path and starting from $0$.

The total full-RSB free energy functional is finally given by:
\begin{equation}
\label{freeFull1}
\Phi=\overline{\int \prod^c_{i=1} d\nu\big(W_i^{(0)}\,\big)\,\,\,\, \phi^{\text{(v)}}\big(0,\bm{W}^{(0)}\,\big)}-\frac{c}{2}\overline{\int d\nu\big(W_1^{(0)}\big)d\nu\big(W_2^{(0)}\big) \phi^{\text{(e)}}\big(\,0,\,W_1^{(0)},W_2^{(0)}\,\big)}\,,
\end{equation}
where the overline $\overline{\,\cdot\,}$ represents the average over the random couplings.

The representation \eqref{GibbsFunctional4}, is an extension to the actual model of the Chen-Auffinger representation of the Parisi functional, defined for the $SK$ model. Note that, in the Chen-Auffinger representation, the process $L^{\text{(e/v)}}$ is a Markov process \cite{ChenAuf}.

The free energy \eqref{freeFull1} actually provides a mathematical representation of the full-RSB free energy, then it may represent an interesting the starting point for further qualitative analysis about replica symmetry breaking on sparse graphs.

For the quantitative evaluation of the free energy, we reduce the present computation to It\^o processes, for which the auxiliary variational problem can be explicitly solved. 

\subsection{The full-RSB equations}
\label{fullRSBsec}
In the preceding subsection, the $full-$RSB ansatz is presented using a variational representation based on continuous martingales on a whatever probability space. This formalism allows deriving a well defined “continuous version” of the variational representation \eqref{GibbsFunctional} of the edge and vertex contributions. However, the generality of the probability space does not enable practical calculations. 

In this subsection, we consider the cavity field and the auxiliary martingales $L^{\text{(e/v)}}$ are explicit functionals of a vectorial Brownian motion on $[0,1]$, with a random starting point. 

We set $\Omega:=C([0,1],\mathbb{R})$ to be the the space of continuous functions $\omega:[0,1]\to \mathbb{R}$ and let $\mathbb{W}_{\nu}$ be the probability measure such as the coordinate map process $W(\omega):=\{W(x,\omega)=\omega(x);\,x\in[0,1]\}$, together with its natural filtration, is a Brownian motion, starting from a random normal distributed point $\omega(0)$ \cite{YoRev}. Such Brownian motion will be simply indicated by $\omega$. The normal distribution will be indicated by $\nu$, according to the notation described at the end of the subsection \eqref{finitersbpara}.

The cavity field functional $h$ is a real valued (Brownian) functional $h:C([0,1],\mathbb{R})\to \mathbb{R}$, measurable with respect to the completion of the $\sigma-$algebra generated by the Brownian motion, and the martingales $L^{\text{(v)}}$ and $L^{\text{(e)}}$ are adapted to the natural completed filtration of the vectorial Brownian motion with $2$ or $c$ components, respectively for the edge and the vertex contribution.

This setting allows several explicit mathematical manipulations. The auxiliary variational problem is solved in the first paragraph, leading to an explicit formula of the variational free energy as a functional of a suitable order parameter. The mean-field equations of the model are then derived in the second paragraph from the first variation of the free energy with respect to the order parameters.

\paragraph{Solution of the auxiliary variational problem.} The martingale representation theorem \cite{Clark}\cite{ItoRep}, assures that the random variables $\bm{h}$ and the matingale $L^{\text{(e/v)}}$ can be represented as It\^o integrals \cite{Ito}:
\begin{equation} 
\label{uRep}
h_i\big(\omega_i\big)=h^{(0)}_i(\omega_i(0))+\int^1_0 \Xi (\,x', \,\omega_{i}\,)d\omega_i(x')
\end{equation}
\begin{equation}
\label{RRep}
L^{\text{(e/v)}}(x,\bm{\omega}\,)=\sum^{(2/c)}_{j=1}\int^{x}_0x'\, r_j^{\text{(e/v)}}(x',\,\bm{\omega}\,)d\omega_j(x')=\int^{x}_0 x' \,\bm{r}^{\text{(e/v)}}(x', \bm{\omega}\,)\cdot d\bm{\omega}(x')\,,
\end{equation}
where $\bm{\omega}$ is the vectorial Brownian motion, with $2$ or $c$ independent components. The subscripts $i$ and $j$ indicate the single components of the vectorial processes $\bm{r}^{\text{(e/v)}}$, $\bm{h}$ and $\bm{\omega}$. The process $\Xi$ and the components of the vector $\bm{r}^{\text{(e/v)}}$ are locally square integrable and adapted processes and the integrals are It\^o stochastic integrals \cite{YoRev}. The function $h^{(0)}_i:\mathbb{R}\to\mathbb{R}$ is an even measurable function. 

Since the cavity fields are independent and identical distributed for each site $i$, the process $\Xi$ is a scalar quantity and independent on $i$.

The processes $\bm{r}^{\text{(e)}}$ and $\bm{r}^{\text{(v)}}$ are determined by the edge and vertex auxiliary variational problems, respectively. 
 
The shorthand notation $(x,\bm{\omega})$ ( or $(x,\omega)$), after a symbol indicating a stochastic process, denotes that such process depends $x$ and has a non-anticipating functional dependence on the Brownian motion $\bm{\omega}$, i.e. it depends on the realization of the Brownian motion $\bm{\omega}(x')$ (or $\omega(x')$ ) at each time $0\leq x' \leq x$:
\begin{gather}
\Xi(x,\omega)=\Xi(x,\{\omega(x');0\leq x'\leq x\}),\\
\bm{r}^{\text{(e/v)}}(x,\bm{\omega})=\bm{r}^{\text{(e/v)}}(x,\{\bm{\omega}(x');0\leq x'\leq x\}),\\
L^{\text{(e/v)}}(x,\bm{\omega})=L^{\text{(e/v)}}(x,\{\bm{\omega}(x');0\leq x'\leq x\}).
\end{gather}

Obviously, the functionals $\Xi$ and $\bm{r}^{\text{(e/v)}}$ must be assumed to be regular enough to ensure the functional \eqref{freeFull1} to be bounded.

In the next, it is also assumed that such processes are Malliavin-derivable \cite{Malliavin}. 

Let $\mathfrak{D}^{\text{(e)}}_{[0,1]}$, $\mathfrak{D}^{\text{(v)}}_{[0,1]}$, $\mathfrak{D}_{[0,1]}$ and $\mathfrak{D}^{(0)}$ be the proper spaces of the processes $\bm{r}^{\text{(e)}}$, $\bm{r}^{\text{(v)}}$ and $\Xi$ and the function $h^{(0)}$, respectively.

Under this framework, let $\psi^{\text{(e/v)}}:\mathfrak{D}^{\text{(e/v)}}_{[0,1]}\times\mathfrak{D}_{[0,1]}\times \mathfrak{D}^{\text{(v)}}_{[0,1]}\to \mathbb{R}$ be the full-RSB auxiliary variational functionals for the edge and vertex contributions, given by:
\begin{multline}
\label{funcAux}
\psi^{\text{(e/v)}}\big[\,\bm{r}^{\text{(e/v)}},h^{(0)},\Xi\,\big](\,\bm{\omega}{(0)}\,)=\mathbb{E}_{(\mathbb{W}_{\nu})^{\otimes(2/c)}}\left[\mathcal{E}\left(\int^1_0 x\,\bm{r}^{\text{(e/v)}}\big(x\,, \,\bm{\omega}\big) \cdot d\bm{\omega}(x)\right)\phi^{\text{(e/v)}} ( \,1,\bm{\omega}\,)\bigg| \,\{\bm{\omega}(0)\}\, \right]\\
-\frac{1}{2}\mathbb{E}_{(\mathbb{W}_{\nu})^{\otimes(2/c)}}\left[\int^1_0 dx\,x\, \,\mathcal{E}\left(\int^x_0 x'\,\bm{r}^{\text{(e/v)}}(x', \,\bm{\omega}) \cdot d\bm{\omega}(x')\right)\,\left \|\,\bm{r}^{\text{(e/v)}}(\,x, \,\bm{\omega}\,) \,\right\|^2\,\,\bigg| \,\{\bm{\omega}(0)\}\,\right]\,,
\end{multline}
where $\mathbb{E}_{\mathbb{W}^{\otimes(2/c)}}\left[\cdot|\{\bm{\omega}(0)\}\right]$ is the expectation value with respect to the vectorial Brownian motion $\bm{\omega}$, conditionally to a fixed realization of the starting point $\bm{\omega}(0)$, and
\begin{gather}
\phi ^{\text{(e)}}\left(\,1,\,\bm{\omega}\,\right)=\log \,\Delta^{\text{(e)}} \left(\,h^{(0)}(\omega_i(0))+\int^1_0 \Xi (\,x', \,\omega_{1}\,)d\omega_1(x'),\,h^{(0)}(\omega_2(0))+\int^1_0 \Xi (\,x', \,\omega_{2}\,)d\omega_2(x') \,\right)\,,\\
\phi ^{\text{(v)}}\left(\,1,\,\bm{\omega}\,\right)=\log \,\Delta^{\text{(v)}} \left(\,h^{(0)}_1(\omega_i(0))+\int^1_0 \Xi (\,x', \,\omega_{1}\,)d\omega_1(x'),\,\cdots\,,h^{(0)}(\omega_c(0))+\int^1_0 \Xi (\,x', \,\omega_{c}\,)d\omega_c(x') \,\right).
\end{gather}
 The auxiliary variational representation \eqref{GibbsFunctional4} is given by

\begin{equation}
\label{GibbsTimeFunctional2}
\phi^{\text{(e/v)}}(\,0,\bm{\omega}{(0)}\,)=
\max_{\bm{r}^{\text{(e/v)}}\in \mathfrak{D}^{\text{(e/v)}}_{[0,1]}} \psi^{\text{(e/v)}}\big[\,\bm{r}^{\text{(e/v)}},h^{(0)},\Xi\,\big](\,\bm{\omega}{(0)}\,)\,,
\end{equation}

The maximum in \eqref{GibbsTimeFunctional2} must be obtained by taking the function $h^{(0)}$ and the process $\Xi$ and the random variable $\bm{\omega}(0)$ fixed.

By Cameron-Martin/Girsanov (CMG) theorem \cite{CameronMartin,Girsanov}, the Doléans-Dade exponential in the expectation values can be reabsorbed in a proper change of probability measure, so the representation \eqref{GibbsTimeFunctional2} actually recovers a Chen-Auffinger-like variational representation. Change of measure, however, acts in a non-trivial way on the process $\Xi$, so, for the actual problem, we avoid the CMG transformation and just deal with the representation \eqref{GibbsTimeFunctional2}.

The maximum of the functional \eqref{funcAux} is obtained by imposing the stationary condition with respect to small perturbations of the process $\bm{r}^{\text{(e/v)}}$. For a given a fixed value $\bm{\omega}(0)\in \mathbb{R}^{(2/c)}$, the functional derivative over the process $\bm{r}^{\text{(e/v)}}$ must be evaluated for all $\,x\in[0,1]$ and $\bm{\omega}\in \big(\,C([0,1],\mathbb{R})\,\big)^{(2,c)}$ starting from $\bm{\omega}(0)$, and put to $0$:
\begin{equation}
\label{stationary}
\frac{\delta \psi^{\text{(e/v)}}\big[\,\bm{r}^{\text{(e/v)}},h^{(0)},\Xi\,\big](\,\bm{\omega}(0)\,)}{\delta r_i^{\text{(e/v)}}(\,x,\bm{\omega}\,)} =0
\end{equation}
The maximum is then attained by the adapted process $\bm{r}^{*\text{(e/v)}}$ that is solution of the following equation:
\begin{equation}
\label{solAux}
r_i^{*(e/v)}(x,\,\bm{\omega})=\mathbb{E}_{(\mathbb{W}_{\nu})^{\otimes(2/c)}}\left[\mathcal{E}\left(\int^1_x x'\,\bm{r}^{*\text{(e/v)}}\big(x', \,\bm{\omega}\big) \cdot d\bm{\omega}(x')\right)\,\text{D}_{i,x}\left(\phi^{\text{(e/v)}}( \,\bm{\omega}\,)\,\right)\,\Big|\{\bm{\omega}{(x')};0\leq x'< x\}\right]\,,
\end{equation}
where $\text{D}_{i,x}\left(\phi_1^{\text{(e/v)}}(\bm{\omega})\,\right)$ is the Malliavin derivative \cite{Malliavin}at time $x$ with respect to the Brownian motion component $\omega_i$ of the Wiener functional $\phi^{\text{(e/v)}}(1,\,\Cdot)$ and $\mathbb{E}_{\mathbb{W}^{\otimes(2/c)}}\left[\cdot|\{\bm{\omega}{(x')};0\leq x'< x\}\right]$ the expectation value conditionally to a fixed realization of the vectorial Brownian motion $\bm{\omega}$ before the time $x$. In the Chen-Auffinger representation, the auxiliary variational functional is convex. We guess that, also in this case,  the functional $\eqref{funcAux}$ has a unique minimizer.

\paragraph{Equilibrium free energy and mean-field equations.} The true variational free energy function $\Phi$ is derived from \eqref{freeFull1}, by substituting the solution of the equation\eqref{solAux}in the representation \eqref{GibbsTimeFunctional2} of the edge and vertex contributions. In such a way, we get a variational functional depending explicitly on the process $\Xi$ and the function $h^{(0)}$. Note that the functional $\Phi:\,\mathfrak{D}^{(0)}\times\mathfrak{D}_{[0,1]}\to \mathbb{R}$ is actually a functional of Brownian functionals. 

The function $h^{(0)}$ and the process $\Xi$ determine uniquely the cavity field functional $h$, so they encode entirely the cavity fields distributions inside the pure states and constitute the true full-RSB order parameters. 

Using the following notation
\begin{equation}
\label{g}
g^{\text{(e/v)}}(x,\bm{\omega}\,)=\mathcal{E}\left(\int^x_0 x'\,\bm{r}^{*\text{(e/v)}}(x' ,\bm{\omega}) \cdot d\bm{\omega}(x')\right)\,,
\end{equation}
the free energy functional is
\begin{multline}
\label{freeFullLast}
\Phi=\Phi[h^{(0)},\Xi]=\overline{\mathbb{E}_{(\mathbb{W}_{\nu})^{\otimes c}}\left[ g^{\text{(v)}}(1, \bm{\omega})\phi^{\text{(v)}}(1,\bm{\omega})\right]}
-\frac{1}{2} \overline{\mathbb{E}_{(\mathbb{W}_{\nu})^{\otimes c}}\bigg[\int^1_0 dx\,\,x\,g^{\text{(v)}}(x, \,\bm{\omega})\,\left\|\,\bm{r}^{*\text{(v)}}(x,\bm{\omega})\,\right\|^2\,\,\bigg]}\\
-\frac{c}{2}\left(\overline{\mathbb{E}_{(\mathbb{W}_{\nu})^{\otimes 2}}\left[g^{\text{(e)}}(1,\omega_1,\omega_2) \phi^{\text{(e)}}(1,\omega_1,\omega_2)\,\right]}
-\frac{1}{2}\overline{ \mathbb{E}_{(\mathbb{W}_{\nu})^{\otimes 2}}\bigg[\int^1_0 dx\,\,x\,g^{\text{(e)}}(x,\omega_1,\omega_2)\,\left\|\,\bm{r}^{*\text{(e)}}(\,x,\omega_1,\omega_2)\,\right\|^2\,\,\bigg]}\right)\,,
\end{multline}

where the symbol $\mathbb{E}_{(\mathbb{W}_{\nu})^{\otimes (2/c)}}[\,\cdot\,]$ is the expectation value with respect to the whole $2/c$-components vectorial Brownian motion $\bm{\omega}$.

The equilibrium free energy $F$ is finally derived by the physical variational problem, through the minimization of the functional $\Phi$ with respect to the process $\Xi$ and the function $h^{(0)}$:
\begin{equation}
F=-\beta \min_{\substack{h^{(0)}\in\mathfrak{D}^{(0)}\\ \Xi \in \mathfrak{D}_{[0,1]}}} \Phi[h^{(0)},\Xi]\,.
\end{equation}

The physical variational problem can be addressed by imposing the stationary condition with respect to the order parameters, by putting to $0$ the first variation of the functional $\Phi$ with respect to the process $\Xi$ and the function $h^{(0)}$. 

Although the equation \eqref{solAux} has not been solved explicitly, the first variation with respect to the physical order parameters can be performed easily, thank to the stationary criterion \eqref{stationary}, in such a way:
\begin{multline}
\begin{aligned}
&\frac{\delta \psi^{\text{(e/v)}}\big[\,\bm{r}^{\text{*(e/v)}}[h^{(0)},\Xi],h^{(0)},\Xi\big](\,\bm{\omega}(0)\,)}{\delta h^{(0)}(\,u(0)\,)} =\\
&\Bigg(\frac{\delta \psi^{\text{(e/v)}}\big[\,\bm{r}^{\text{(e/v)}},h^{(0)},\Xi\,\big](\,\bm{\omega}(0)\,)}{\delta h^{(0)}(\,u(0)\,)} +\\
&\sum^{(2/c)}_{i=1}\int^1_0dx'\,\mathbb{E}_{(\mathbb{W}_{\nu})^{\otimes c}}\left[\,\frac{\delta r_i^{\text{*(e/v)}}[h^{(0)},\Xi](\,x',\bm{\omega}\,)}{\delta h^{(0)}(\,u(0)\,)} \frac{\delta \psi^{\text{(e/v)}}\big[\bm{r}^{\text{(e/v)}},h^{(0)},\Xi\,\big](\,\bm{\omega}(0)\,)}{\delta r_i^{\text{(e/v)}}(\,x',\bm{\omega}\,)} \Bigg| \,\{\bm{\omega}(0)\}\, \,\right]\Bigg)\Bigg|_{\bm{r}^{\text{(e/v)}}=\bm{r}^{*\text{(e/v)}}[h^{(0)},\Xi] }
\end{aligned}\\
=\Bigg(\frac{\delta \psi^{\text{(e/v)}}\big[\,\bm{r}^{\text{(e/v)}},h^{(0)},\Xi\,\big](\,\bm{\omega}(0)\,)}{\delta h^{(0)}(\,u(0)\,)} \Bigg)\Bigg|_{\bm{r}^{\text{(e/v)}}=\bm{r}^{\text{*(e/v)}}[h^{(0)},\Xi]}
\end{multline}
and
\begin{multline}
\begin{aligned}
&\frac{\delta \psi^{\text{(e/v)}}\big[\,\bm{r}^{\text{*(e/v)}}[h^{(0)},\Xi],h^{(0)},\Xi\,\big](\,\bm{\omega}(0)\,)}{\delta \Xi(\,x, u\,)} =\\
&\Bigg(\frac{\delta \psi^{\text{(e/v)}}\big[\,\bm{r}^{\text{(e/v)}},h^{(0)},\Xi\,\big](\,\bm{\omega}(0)\,)}{\delta \Xi(\,x, u\,)} +\\
&\sum^{(2/c)}_{i=1}\int^1_xdx'\,\mathbb{E}_{(\mathbb{W}_{\nu})^{\otimes c}}\left[\,\frac{\delta r_i^{\text{*(e/v)}}[h^{(0)},\Xi](\,x',\bm{\omega}\,)}{\delta \Xi(\,x, u\,)} \frac{\delta \psi^{\text{(e/v)}}\big[\bm{r}^{\text{(e/v)}},h^{(0)},\Xi\,\big](\,\bm{\omega}(0)\,)}{\delta r_i^{\text{(e/v)}}(\,x',\bm{\omega}\,)} \Bigg| \,\{\bm{\omega}(0)\}\, \,\right]\Bigg)\Bigg|_{\bm{r}^{\text{(e/v)}}=\bm{r}^{*\text{(e/v)}}[h^{(0)},\Xi] }
\end{aligned}\\
=\Bigg(\frac{\delta \psi^{\text{(e/v)}}\big[\,\bm{r}^{\text{(e/v)}},h^{(0)},\Xi\,\big](\,\bm{\omega}(0)\,)}{\delta \Xi(\,x, u\,)} \Bigg)\Bigg|_{\bm{r}^{\text{(e/v)}}=\bm{r}^{\text{*(e/v)}}[h^{(0)},\Xi]}
\end{multline}
where the dependence of $\bm{r}^{\text{*(e/v)}}$ is written explicitly.

This is another important goal that motivates the auxiliary variational representation approach presented in this paper.

The stationary condition leads to a system of well-defined coupled functional mean-field equations for the order parameters $\Xi$ and $h^{(0)}$.

Let $\bm{m}^{\text{(e)}}: C([0,1],\mathbb{R})\to \mathbb{R}^{2}$ and $\bm{m}^{\text{(v)}}: C([0,1],\mathbb{R})\to \mathbb{R}^{c}$ be the vectorial Brownian functionals 

\begin{equation}
\bm{m}^{\text{(e/v)}}(1,\bm{\omega})=\bm{M}^{\text{(e/v)}}\left(\,\,h^{(0)}_1(\omega_i(0))+\int^1_0 \Xi (\,x', \,\omega_{1}\,)d\omega_1(x'),\,\cdots\,,h^{(0)}(\omega_{\,2/c}(0))+\int^1_0 \Xi (\,x', \,\omega_{\,2/c}\,)d\omega_{\,2/c}(x')\,\right)\,,
\end{equation}
with
\begin{equation}
\bm{M}^{\text{(e/v)}}\big(\,\bm{y}\,\big)=\nabla\log \,\Delta \left(\,\bm{y}\right)\,, \quad \text{with}\,\,y\in \mathbb{R}^{(2/c)},
\end{equation}
where the symbol $\nabla$ denotes the gradient over the variables $\bm{y}$.

After some manipulation, the first variation with respect to the function $h^{(0)}$ gives the equation
\begin{multline}
\label{Eqh}
\frac{1}{2}\sum^2_{i=1}\overline{\mathbb{E}_{(\mathbb{W}_{\nu})^{\otimes 2}}\left[\,g^{\text{(e)}}\big(1,\omega_1,\omega_2\big) m_i^{\text{(e)}}(1,\omega_1,\omega_2)\,\delta(\omega_i(0)-u )\,\right]}\,=\\
\frac{1}{c}\sum^c_{i=1}\overline{\mathbb{E}_{(\mathbb{W}_{\nu})^{\otimes c}}\left[\,g^{\text{(v)}}(1,\bm{\omega}) m_i^{\text{(v)}}(1,\bm{\omega})\,\delta(\omega_i(0)-u) \,\right]}\,,\quad \forall u\in \mathbb{R}
\end{multline}
and the first variation on the process $\Xi$ gives the equation
\begin{multline}
\label{processEQ}
\frac{1}{2}\sum^2_{i=1}\overline{\mathbb{E}_{(\mathbb{W}_{\nu})^{\otimes 2}}\left[\,\delta_{[0,x]}\left[u(\Cdot)-\omega_i(\Cdot)\,\right]\,D_{i,x}\left( g^{\text{(e)}}\big(1,\omega_1,\omega_2\big) m_i^{\text{(e)}}(1,\omega_1,\omega_2)\right)\,\right]}\,=\\
\frac{1}{c}\sum^c_{i=1}\overline{\mathbb{E}_{(\nu\times\mathbb{W}_{\nu})^{\otimes c}}\left[\delta_{[0,x]}\left[u(\Cdot)-\omega_i(\Cdot)\,\right]\,D_{i,x}\left(g^{\text{(v)}}(1,\bm{\omega}) m_i^{\text{(v)}}(1,\bm{\omega}) \,\right)\,\right]}\,,\quad \forall\,u\in C([0,1],\mathbb{R})\,\,\text{and}\,\,x\in[0,1]
\end{multline}
where the symbol $\delta(\omega_i(0)-u )$ denote the Dirac delta distribution and the symbol $\delta_{[0,x]}[\,\cdot\,]$ is the functional Dirac delta over the realization of the considered process $\omega_i$ up to the time $x$ and.

Calling $\Xi^*$ and $h^{*(0)}$ the solutions of the previous equations, the equilibrium $F$ is equal to $F=-\beta \Phi[h^{*(0)},\Xi^*]$.

\section{Summary and conclusion}
In this paper, the first non-ambiguous description of the full-RSB formalism for the Ising spin glass on random regular graphs is obtained.

The order parameter is a stochastic functional, related to the distribution of the cavity fields populations at each site \eqref{AustinRep}, in contrast with the order parameter in the Mézard-Parisi ansatz which involves a hierarchical tower of distributions of distributions, hard to extend to the full-RSB limit.

Since the order parameter is a functional, the extension of the discrete RSB scheme to the continuous case cannot be achieved simply through a (generalized) Parisi-like partial differential equation.

This problem has been overcome by formalizing the ideas suggested by G. Parisi, in \cite{ParisiMarginal}, in a proper new stochastic variational approach, that provides a powerful mathematical tool to study such class of models.

The approach proposed in this paper allows deriving a well-defined free energy functional from which the self-consistency mean-field equations can be easily derived.

We get, then, a complete definition of the full-RSB problem for the Ising spin glass on random regular graph, with a given free energy, depending on certain order parameters, and the equation for the order parameters.

We guess that the solution of the presented mean-field equations provides the right equilibrium free energy, but a rigorous proof is still missing. It is worth noting, however, that the mathematical formalization of the full-RSB scheme, proposed in that work, is a necessary groundwork for a rigorous analysis of the RSB phenomenon-beyond the fully-connected models.

Unfortunately, the mean-field equations are very difficult to solve. Since the order parameter is a functional of a Brownian motion, we guess that the self-consistency equation may be solved by a population dynamics over populations of Brownian motion paths.

The quantitative evaluation of the free energy and a deeper analysis of the physics properties will be investigated in next works.

\section*{Acknowledgements}
I am grateful to Giorgio Parisi and Simone Franchini for interesting discussions. 
\appendix
\section {The 1 RSB free energy functional}
\label{appendix}
In this appendix, we show that the replica symmetry ansatzes described in the subsection \eqref{finitersbpara} reproduce the Mézard-Parisi 1-RSB cavity method.

Let us define the functions
\begin{equation}
F^{\text{(v)}}(J_{0,1},\cdots,J_{0,c},h_1,\cdots,h_c\,)=\log\Delta^{\text{(v)}}(J_{0,1},\cdots,J_{0,c},h_1,\cdots,h_c\,),\\
\end{equation}
and
\begin{equation}
F^{\text{(e)}}(J_{1,2},h_1,h_2\,)=\Delta^{\text{(e)}}(J_{1,2},h_{1},h_{2})\,.
\end{equation}

Putting $r=1$, the cavity field functional \eqref{AustinRep} is a measurable function of two independent normal random variables $W^{(0)}$ and $W^{(1)}$:
\begin{equation}
\{W^{(0)},W^{(1)}\}\mapsto h\big(\,W^{(0)},W^{(1)}\,\big).
\end{equation}

The edge and vertex contributions to the free energy are given by a single iteration of the iterative rule \eqref{start1}:
\begin{multline}
\label{start2}
\phi^{\text{(e/v)}}_{0}\big(\bm{J},\,\bm{W}^{(0)}\,\big)=\frac{1}{x_{1}}\log \mathbb{E}_{\mathbb{W}_1^{\otimes (2/c)}}\left[\,\exp\left(\,x_{1} \phi^{\text{(e/v)}}_{1}\big(\bm{J},\bm{W}^{(0)},\,\bm{W}^{(1)}\,\big)\,\right)\,\Big|\bm{W}^{(0)}\right]=\\
\frac{1}{x_{1}}\log\left(\int \left(\prod^{(2/c)}_{i=1} d\nu\big(W^{(1)}_{i}\big)\,\right) e^{\,x_{1}\, F^{\text{(e/v)}}\left(\bm{J},\bm{h}\big(\,\bm{W}^{(0)},\,\bm{W}^{(1)}\,\big)\,\,\right)\,}\,\,\right)\,,
\end{multline}
where $\nu$ is the normal distribution.

The free energy functional is finally given by
\begin{multline}
\label{freeFull1RSB}
\Phi=\overline{\int \prod^c_{i=1} d\nu\big(W_i^{(0)}\,\big)\,\,\frac{1}{x_{1}}\log\left(\int\left(\prod^{c}_{i=1} d\nu\big(W^{(1)}_{i}\big)\,\right) e^{\,x_{1}\, F^{\text{(v)}}\left(J_{0,1},\cdots,J_{0,c},\,h_1\big(\,W_1^{(0)},\,W_1^{(1)}\,\big),\cdots\,h_c\big(\,W_c^{(0)},\,W_c^{(1)}\,\big)\,\right)\,}\,\,\right)}\\-\frac{c}{2}\overline{\int d\nu\big(W_1^{(0)}\big)d\nu\big(W_2^{(0)}\big) \frac{1}{x_{1}}\log\left(\int_{\mathbb{R}^{(2/c)}}d\nu\big(W_1^{(1)}\big)d\nu\big(W_2^{(1)}\big)\,e^{\,x_{1}\, F^{\text{(e)}}\left(J_{1,2},h_1\big(\,W_1^{(0)},\,W_1^{(1)}\,\big),\,h_2\big(\,W_2^{(0)},\,W_2^{(1)}\,\big)\,\right)\,}\,\,\right)}\,.
\end{multline}
Now, let us define the probability density distribution of the cavity field $h$, conditionally to a fixed value for the random variable $W^{(0)}$:
\begin{equation}
\label{Mypi}
\widehat{\pi}\big(y\big|W^{(0)}\,\big)=\mathbb{E}_{\mathbb{W}_1^{\otimes (2/c)}}\left[\delta\big(\,y-h\big(\,W^{(0)},W^{(1)}\,\big)\,\big)\Big|W^{(0)}\right]=\int d\nu\big(W^{(1)}\big)\delta\big(\,h\big(\,W^{(0)},W^{(1)}\,\big)-y\,\big)\,,\quad y\in \mathbb{R}
\end{equation}
For each value of $y\in \mathbb{R}$ fixed, the quantity $\widehat{\pi}(y|\,\Cdot\,)$ is a positive random variable, since it depends on $W^{(0)}$. Then, we can define the probability density distribution of the random probability density distribtion $\widehat{\pi}:=\big\{\widehat{\pi}\big(y\big|\,\Cdot\,\big);\,y\in \mathbb{R}\,\big\}$ in such a way:
\begin{equation}
\label{MyPi}
\Pi[\,\pi\,]=\mathbb{E}_{\mathbb{W}_0^{\otimes (2/c)}}\left[\delta\big[\,\pi\big(\,\Cdot\,\big)-\widehat{\pi}\big(\,\Cdot\,\big|W^{(0)}\,\big)\,\big]\,\right]=\int d\nu\big(W^{(0)}\big)\delta\big[\,\pi\big(\,\Cdot\,\big)-\widehat{\pi}\big(\,\Cdot\,\big|W^{(0)}\,\big)\,\big]\,,
\end{equation}
where $\delta[\,\Cdot\,]$ is the functional Dirac delta.
Substituting the equation \eqref{Mypi} and \eqref{MyPi} in\eqref{freeFull1RSB}, one get
\begin{multline}
\label{ParMez}
x_1 \Phi=x_1 \Phi\big[\Pi,\pi,x_1\big]=\overline{\int \left(\prod^c_{i=1} d[\pi_i]\Pi[\,\pi_i\,]\right)\,\,\log\left(\int\left(\prod^{c}_{i=1} dy \,\pi\big(y\big)\,\right) e^{\,x_{1}\, F^{\text{(v)}}\left(J_{0,1},\cdots,J_{0,c},\,y_1,\cdots\,y_c\,\big)\,\right)\,}\,\,\right)}\\-\frac{c}{2}\overline{\int d[\pi_1]\Pi[\,\pi_1\,]\,\,d[\pi_2]\Pi[\,\pi_2\,] \log\left(\int_{\mathbb{R}^{(2/c)}}\,dy_1\,\pi_i(y_1)\,dy_2\,\pi_i(y_2)\,e^{\,x_{1}\, F^{\text{(e)}}\left(J_{1,2},y_1,\,y_2\,\right)\,}\,\,\right)}\,,
\end{multline}
where the symbol $\int d[\pi]$ represents the functional integral.

The variational free energy functional \eqref{ParMez} is equivalent to a finite temperature version of the 1-RSB variational free energy presented in appendix B of \cite{ParMezRRG4}. The functional $\Pi$ is the 1-RSB cavity method order parameter and $x_1$ is the Parisi 1-RSB parameter.



\end{document}